\documentclass[accepted]{uai2025} 

\usepackage[american]{babel}

\usepackage{natbib} 
\usepackage{mathtools} 
\usepackage{booktabs} 
\usepackage{tikz} 

\usepackage{amsmath,amsfonts,bm}

\def\eqref#1{equation~\ref{#1}}

\def\1{\bm{1}}

\def\vone{{\bm{1}}}

\def\vz{{\bm{z}}}

\def\mA{{\bm{A}}}
\def\mB{{\bm{B}}}

\def\mF{{\bm{F}}}

\def\mP{{\bm{P}}}
\def\mQ{{\bm{Q}}}

\def\mU{{\bm{U}}}
\def\mV{{\bm{V}}}

\def\mX{{\bm{X}}}
\def\mY{{\bm{Y}}}

\def\mSigma{{\bm{\Sigma}}}

\DeclareMathAlphabet{\mathsfit}{\encodingdefault}{\sfdefault}{m}{sl}
\SetMathAlphabet{\mathsfit}{bold}{\encodingdefault}{\sfdefault}{bx}{n}

\def\gA{{\mathcal{A}}}

\def\sR{{\mathbb{R}}}

\newcommand{\R}{\mathbb{R}}

\usepackage{hyperref}
\usepackage{url}
\def\bmgA{{\boldsymbol{\gA}}}
\def\vtau{{\bm{\tau}}}
\usepackage{algorithm}
\usepackage{algpseudocode} 
\usepackage{graphicx}
\usepackage{subcaption}
\usepackage{bbm}
\usepackage{amsmath}
\usepackage{extpfeil}
\usepackage{amsthm}
\usepackage{amsmath}
\usepackage{natbib}
\usepackage{bm}
\usepackage{mathrsfs}
\newtheorem{lemma}{Lemma}
\usepackage{tikz}
\usetikzlibrary{arrows.meta, calc, chains, quotes, positioning, shapes.geometric}
\usepackage{placeins}

\title{Valid Bootstraps for Network Embeddings with Applications to Network Visualisation}

\author[1]{\href{mailto:<emerald.dilworth@bristol.ac.uk>?Subject=Valid Bootstraps for Network Embeddings with Applications to Network Visualisation}{Emerald Dilworth}{}}
\author[1]{Ed Davis}
\author[1]{Daniel J. Lawson}
\affil[1]{%
    School of Mathematics\\
    University of Bristol\\
    Bristol, UK
}
  
\begin{document}
\maketitle

\begin{abstract}
Quantifying uncertainty in networks is an important step in modelling relationships and interactions between entities. We consider the challenge of bootstrapping an inhomogeneous random graph when only a single observation of the network is made and the underlying data generating function is unknown. 
We address this problem by considering embeddings of the observed and bootstrapped network that are statistically indistinguishable. 
We utilise an exchangeable network test that can empirically validate bootstrap samples generated by any method. Existing methods fail this test, so we propose a principled, distribution-free network bootstrap using k-nearest neighbour smoothing, that can pass this exchangeable network test in many synthetic and real-data scenarios. We demonstrate the utility of this work in combination with the popular data visualisation method t-SNE, where uncertainty estimates from bootstrapping are used to explain whether visible structures represent real statistically sound structures. 
\end{abstract}

\section{INTRODUCTION}\label{section:introduction}

Networks are ubiquitous across many applications including cyber-security \citep{bowman2021towards, he2022illuminati, bilot2023graph}, biology \citep{brain_embedding_node2vec, jumper2021highly}, natural language \citep{word2vec}, and recommendation algorithms \citep{wu2022graph}. Despite the influence of this field, uncertainty quantification for networks remains a challenge.

The bootstrap method introduced by \cite{efron1979bootstrap} provides an estimator for the mean and can be adapted to estimate variance, order statistics, or any other functional of the data for i.i.d. samples \citep{efron1994introduction}. With appropriate quantile corrections, these estimates are provably correct. Handling dependency requires care \citep{kreiss2011bootstrap} and so networks, for which edges are far from independent, require special treatment.
Many networks can be described as following a limiting distribution \citep{lovasz2006limits,borgs2008convergent}, where edges are independent given the node cluster labels. \cite{green2022bootstrapping} apply this to bootstrapping networks  using a graphon-based `histogram bootstrap'. 
This approach is further developed by \cite{zu2024local} for valid local network statistics. 

More generally, a Random Dot Product Graph (RDPG) places nodes in a latent vector space and edges become independent conditional on location, which generates provably correct bootstraps \citep{levin2019bootstrapping}. Adjacency Spectral Embedding (ASE) \citep{hoff2002latent, sussman2012consistentadjacencyspectralembedding} provides strong theoretical asymptotic guarantees \citep{Cape2019TwoToInfinity}. These can be generalised to multipartite networks \citep{RubinDelanchy2022statistical}.
However, inference for these methods relies on ASE, which empirically has lower power relative to other available embedding methods \citep{qin2013regularized, grover2016node2vec}.  Nearest-neighbour methods \citep{stoneConsistentNonparametricRegression1977} also provide asymptotic convergence to a latent position \citep{lianConvergenceFunctionalKnearest2011} and can be applied to nonlinear embeddings such as Node2Vec \citep{grover2016node2vec} and ProNE \citep{Zhang2019Prone}. 

Further, we care about finite-sample properties. To address this, we apply developments in `Unfolding' \citep{Lathauwer2000SVD,gallagher2021spectral} to construct a Bootstrap Exchangeability Test (Section \ref{section:exchangeability}) for whether the embeddings of a bootstrapped network and the observed network are exchangeable, in a finite dataset. This reveals that the nonlinear models outperform linear ones and are needed for practical usage.

Previous work uses count statistics and U-statistics to summarise the performance of network bootstraps \citep{bhattacharyya2015subsampling,epskamp2018estimating,levin2019bootstrapping}. These approaches test that the distribution of local features of networks are conserved under bootstrap, e.g. the counts of triangles. Counting triangles is appropriate for some applications \citep{perez2014community, gao2022clustering}, but these structures can be averaged-out in embeddings. 
Complete U-statistics can be impossible to represent in low dimensions \citep{seshadhri2020impossibility}
, but fortunately they are not required for downstream tasks such as network visualisation \citep{mcinnes2018umap, van2008visualizing}, anomaly detection \citep{akoglu2015graph}, and graph similarity comparisons \citep{koutra2013deltacon}, which are the focus of this paper. For such tasks, even `useful' bootstrap procedures are often dismissed, with no readily available replacements. 
We therefore introduce a suitable notion of validity for embeddings, 
which insists that the observed and bootstrapped network have statistically indistinguishable embedding distributions. 
This can be formalised into a test for whether network bootstraps are exchangeable with the observed network in a joint embedding. The joint embedding is chosen to have the \textit{stability} property, i.e. nodes that have the same connectivity distribution but belong to different networks still receive the same embedding. 
Generating bootstraps that remain valid in an embedding space is crucial for many downstream tasks and is the focus of our contributions. 

\begin{figure*}[ht]
    \centering
    \includegraphics[width=0.95\linewidth]{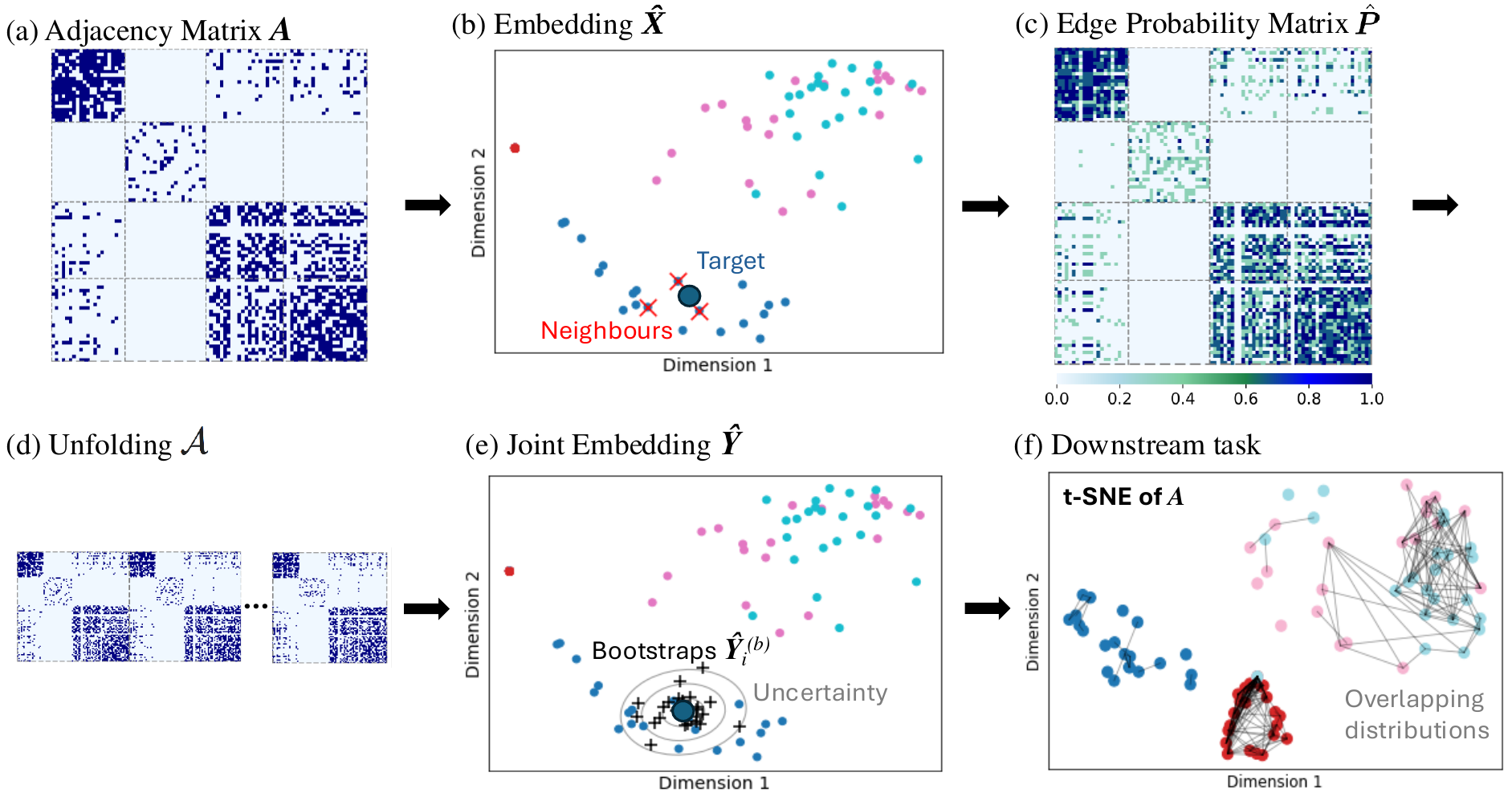}
    \caption{An overview of the proposed graph bootstrap framework. 
    (a): Observe an $n$ node network as an adjacency matrix $\mA \in \{0,1\}^{n \times n}$, where it is assumed that $\mA_{ij} \overset{\text{ind}}{\sim}\text{Bernoulli}(\mP_{ij})$. 
    (b): Embed the network into $d \ll n$ dimensions, e.g. via ASE, to obtain a low-dimension representation of the network $\hat{\mX} \in \mathbb{R}^{n \times d}$. 
    (c): Find the $k$-nearest neighbours for each node and use these as in Algorithm \ref{alg:knn_bootstrap} to estimate the probability matrix as $\hat{\mP} \in [0,1]^{n \times n}$. 
    (d): Generate $B$ bootstrap resamples of the observed network by sampling $\widetilde{\mA}^{(b)}_{ij} \overset{\text{ind}}{\sim}\text{Bernoulli}(\hat{\mP}_{ij})$ for $b=1,...,B$. 
    (e): Use the bootstrap resamples for downstream tasks, such as estimating nodewise variance, as shown. 
    (f): We can link nodes by whether their distributions overlap in an visualisation, here t-SNE.
    } 
    \label{fig:bootstrap_outline}
\end{figure*}

Figure \ref{fig:bootstrap_outline} explains our contributions. (a) The Adjacency matrix $\mA$ is modelled as $\mA_{ij} \overset{\text{ind}}{\sim} \text{Bernoulli}\left(\bm{P}_{ij}\right)$, for $i,j \in \{1,...,n\}$. (b) We then compute an embedding $\hat{\mX}$ of the network which (c) we utilise to form a broad range of estimators for $\bm{P}$; we add k-Nearest Neighbours on nonlinear embeddings to the options. (d) Unfolding the observed and bootstrapped adjacency matrices, $\mA$ and $\tilde{\bm{A}}$, respectively, allows them to be (e) embedded into a shared space, from which downstream tasks of uncertainty quantification can be performed. These include our Bootstrap Exchangeability Test and (f) visualising uncertainty in t-SNE \citep{van2008visualizing} embeddings.

\section{THEORETICAL BACKGROUND}\label{section:theoretical_background}

Our proposed bootstrapping procedure requires a method for embedding a single network. For our proposed method to validate the bootstrap, we need a method for embedding multiple networks. We introduce both types of embedding in this section, as well as the concept of exchangeability, which are core concepts in our main contributions.

\subsection{Embedding Single Networks}
An undirected $n$ node network is represented using a binary and symmetric adjacency matrix $\mA \in \{0, 1\}^{n \times n}$, where $\mA_{ij} = 1$ if an edge exists between nodes $i$ and $j$ and $\mA_{ij} = 0$ otherwise for each $i, j \in \{1, \dots, n\}$. This paper considers networks which can be modelled using a binary inhomogeneous random graph (BIRG) \citep{soderberg2002general}, 
\[\mA_{ij} \overset{\text{ind}}{\sim}\text{Bernoulli}(\mP_{ij}),\]
where $\mP_{ij}$ denotes the probability of an edge 
between nodes $i$ and $j$. For the remainder of this paper, unless otherwise stated, we 
assume that each $\mA$ is drawn from a BIRG.

It is often useful to consider the graph using a low-dimensional representation $\hat{\mX} \in \R^{n \times d}$, referred to as embedding. The embedding dimension $d$ is often chosen such that $d \ll n$. While our contributions are not limited to a single embedding method, we will focus on ASE due to its simplicity. 
For ASE, it is common practice to observe a scree plot of the singular values and find the `elbow' to decide the value of $d$ \citep{ZHU2006918,screePlotElbow}.

\paragraph{Definition 1} (Adjacency spectral embedding) The $d$-dimensional \textit{adjacency spectral embedding} of a network $\mA$ is given by 
\begin{equation*}
    \hat{\mX} = \hat{\mU}_\mA |\hat{\mSigma}_\mA|^{1/2},
\end{equation*}
where 
$\hat{\mSigma}_\mA \in \R^{d \times d}$ is a diagonal matrix containing the $d$ largest eigenvalues of $\mA$ arranged with decreasing magnitude, and $\hat{\mU}_\mA \in \R^{n \times d}$ is a matrix containing, as columns, the corresponding eigenvectors. Here, the $i$-th row of the embedding matrix $\hat{\mX} \in \R^{n \times d}$ is the transposed $d$-dimensional vector representation of the $i$-th row of $\mA$. 
For single-network embeddings, we use the notation $\hat{\mX}$.

\subsection{Embedding Multiple Networks}\label{section:UASE}


To validate a proposed bootstrap, we embed both the observed network and its bootstrap into the same space. For this task, we require a multiple-network embedding. Specifically, we denote the multi-network embedding by $\hat{\mY} = (\hat{\mY}^{(1)}; \dots; \hat{\mY}^{(M)}) \in \mathbb{R}^{Mn \times d}$, which is a row-wise concatenation of the embeddings of $M$ networks. Here, each $\hat{\mY}^{(m)} \in \R^{n\times d}$ represents the embedding of the $m$-th network $\mA^{(m)}$ obtained from the multi-network embedding. 

One method for multi-network embedding is unfolded adjacency spectral embedding (UASE), which embeds a collection of networks into a single embedding space, allowing for comparisons to be made across the networks \citep{jones2020multilayerRDPG}.

\paragraph{Definition 2} (Unfolded adjacency spectral embedding) Let $\bmgA = \left(\mA^{(1)}, \dots, \mA^{(M)} \right) \in \{0, 1\}^{n \times Mn}$ (column concatenation) be the unfolding of a collection of $M$ $n$-node networks. The $d$-dimensional \textit{unfolded adjacency spectral embedding} of this collection is given by
\begin{equation*}
    \hat{\mY} = \hat{\mV}_{\bmgA} \hat{\mSigma}_\bmgA^{1/2},
\end{equation*}
where $\hat{\mU}_\bmgA \hat{\mSigma}_\bmgA \hat{\mV}_\bmgA^T$ is the rank-$d$ truncated singular value decomposition of $\bmgA$, that is, $\hat{\mSigma}_\mA \in \R^{d \times d}$ is a diagonal matrix containing the $d$ largest singular values of $\bmgA$ arranged in decreasing order, and $\hat{\mU}_\mA$, $\hat{\mV}_\mA$ contain the corresponding left and right singular vectors, respectively. Here, $\hat{\mY} = ( \hat{\mY}^{(1)}; \dots; \hat{\mY}^{(M)} ) \in \R^{Mn \times d}$, where the $i$-th row of $\hat{\mY}^{(m)}$ contains the transposed $d$-dimensional vector representation of the $i$-th row of $\mA^{(m)}$ for each $m\in \{1, \dots, M\}$.


\subsection{Across-Network Exchangeability}\label{section:exchangeability}

UASE is far from the only method available for embedding multiple networks, e.g. \citet{levin2017central, chen2020multiple, scheinerman2010modeling, lin2008facetnet}. However, we utilise UASE as unfolding allows for across-network exchangeability \citep{gallagher2021spectral}, which we leverage for our main contributions. 

\paragraph{Definition 3} (Across-network exchangeability) Let $\mP_i$ denote the $i$-th row of a probability matrix $\mP \in [0,1]^{n \times n}$. Let $\hat{\mathbf{Y}}^{(1)}, \dots, \hat{\mathbf{Y}}^{(T)} \in \mathbb{R}^{n \times d}$ be a $d$-dimensional multi-network embedding of a collection of BIRGs with probability matrices $\mP^{(1)}, \dots, \mP^{(M)}$, respectively. This embedding has the property of across-network exchangeability if when $\mP^{(m)}_i = \mP^{(u)}_i$, we have that 
\begin{align*}
    \mathbb{P} \left(\hat{\mY}_i^{(m)} = v_1, \hat{\mY}_i^{(u)} = v_2 \right) = 
    \mathbb{P} \left(\hat{\mY}_i^{(m)} = v_2, \hat{\mY}_i^{(u)} = v_1 \right),
\end{align*}
for any $m, u \in \{1, \dots, M\}$ and any $i \in \{1, \dots, n\}$ \citep{davis2023simple}. In words, this condition states that embeddings $\hat{\mY}_i^{(m)}$ and $\hat{\mY}_i^{(u)}$ are exchangeable.

Alternative embedding methods can be used and have across-network exchangeability by utilising Dilated Unfolded Embedding \citep{davis2023simple}. For simplicity, we mainly focus on using UASE in this work. See Appendix \ref{dilatedEmbeddingNetTest} for details.

\section{CONTRIBUTIONS}

Our main contributions are firstly, a novel method to validate bootstraps based on their exchangeability with the observed network, and secondly, two bootstrapping methods. One has theoretical guarantees in the asymptotic regime, and another improved empirical performance in the finite regime.

\subsection{Bootstrap Validation Procedure}\label{section:bootstrap_procedure_validation}

In the unrealistic case where we know the true probability matrix $\mP$ from which an observed network $\mA$ was drawn, then we could draw a true resample of that network as $\widetilde{\mA}_{ij} \overset{\text{ind}}{\sim} \text{Bernoulli} (\mP_{ij})$ for each $i, j \in \{1, \dots, n\}$. This case represents the best possible bootstrap for our given model assumptions and the entries of $\mA$ and $\widetilde{\mA}$ will be exchangeable. Consequently, a good bootstrap must produce an $\widetilde{\mA}$ with exchangeable entries to that of $\mA$, based on the finite sample evidence. We therefore propose a method of testing for exchangeability between networks.

Directly comparing adjacency matrices entrywise as $n$ grows is both computationally expensive and conceptually problematic without access to the true $\mP$, as any errors in estimation can lead to bootstrap resamples which are not valid. 
To solve this problem, we opt to compare lower-dimensional embeddings of the networks, which can be estimated. 

Our procedure for bootstrap validation considers an observed $n$ node network $\mA$ and one bootstrap $\widetilde{\mA}$. 
We first compute a $d$-dimensional across-network exchangeable embedding $\hat{\mY} = (\hat{\mY}^{(\text{obs})};\widetilde{\mY}) \in \R^{2n \times d}$, where $\hat{\mY}^{(\text{obs})}$ is the embedding of the observed network $\mA$ and $\widetilde{\mY}$ is the embedding of its bootstrap $\widetilde{\mA}$ in the shared space. 

Due to across-network exchangeability, $\hat{\mY}^{(\text{obs})}$ and $\widetilde{\mY}$ will be exchangeable if $\mA$ and $\widetilde{\mA}$ are drawn from the same distribution. Next, we apply the paired displacement test from \citet{davis2023simple} (a permutation test for a pair of networks), whose null hypothesis is that the two adjacency matrices are exchangeable, 
and therefore $\hat{\mY}^{(\text{obs})}_i$ and $\widetilde{\mY}_i$ have the same latent position, for all $i \in \{1,...,n\}$. 
The hypotheses are thus
\[
\begin{aligned}
H_0 &: \hat{\mY}^{(\text{obs})} \text{ and } \widetilde{\mY} \text{ follow the same distribution}, \\
H_1 &: \hat{\mY}^{(\text{obs})} \text{ and } \widetilde{\mY} \text{ do not follow the same distribution}.
\end{aligned}
\]

We exploit this property for the `Bootstrap Exchangeability Test', which tests the validity of a \emph{single bootstrap sample}. Let $\pi_1,...,\pi_R$ be $R$ permutations of $\hat{\mY}$, where each permutation swaps rows $i$ and $i+n$ with probability $\frac{1}{2}$ for $i=1,...,n$. This gives $R$ permuted versions of $\hat{\mY}$ denoted by $\pi_1(\hat{\mY}),...,\pi_r(\hat{\mY})$. 
For each permutation, we calculate a test statistic $t( \pi_r (\hat{\mY}))$ capturing the displacement of the second network from the first, as:
\begin{equation}\label{eq:t_q}
    T_{r+1} = t( \pi_r (\hat{\mY})) 
    = \left|\left| \sum\limits_{i=1}^N \pi_r(\hat{\mY})_{i} - \sum\limits_{i=N+1}^{2N} \pi_r(\hat{\mY})_{i} \right|\right|_2 .
\end{equation}
From these  $R$ test statistics we calculate a $p$-value: 
\begin{equation}\label{eq:p_val}
    \hat{p} = \frac{1}{R+1} \sum\limits_{r=1}^{R+1} \mathbbm{1}\{T_r \ge t_{obs}\} .
\end{equation}
If the $p$-value provides insufficient evidence for the alternate hypothesis, then the bootstrap is deemed exchangeable with the observed, based on the finite sample evidence, and hence the bootstrap is valid. 



A summary of this procedure is displayed in Algorithm \ref{alg:test}, which for real data scenarios is applied to an observed matrix with each of its $B$ bootstrap replicates individually, giving $B$ $p$-values.

In simulated data scenarios we know the true probability matrix $\mP$, which allows a more robust verification of bootstrap validity by conditioning on $\mP$ rather than $\mA$. 
We therefore take $M$ adjacency matrices draws $\mA^{(1)},...,\mA^{(M)}$  from the \textit{known} probability matrix $\mP$. For each matrix, we draw one bootstrap resample given by $\widetilde{\mA}^{(1)},...,\widetilde{\mA}^{(M)}$, respectively, and apply Algorithm \ref{alg:test} to each pair. 
\begin{figure}[ht]
\begin{center}
\includegraphics[width=0.62\linewidth]{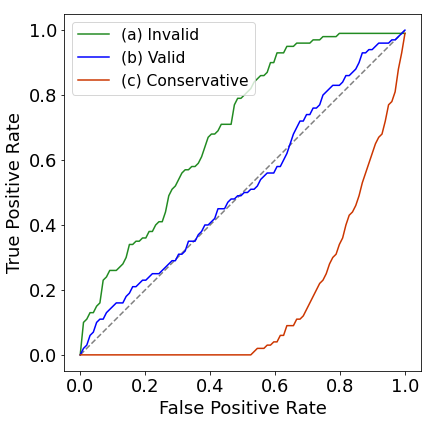}
\end{center}
\caption{Example QQ-plots. (a): An \textbf{invalid resample} has too many small $p$-values and occurs if the resampled data are centred too far from the true latent position, when compared with the embedding of the observed data, which would lead to overestimation of the variance. 
(b): A \textbf{valid resample} has uniformly distributed $p$-values and the bootstrap resamples are exchangeable with the observed - we cannot tell which data come from which network. 
(c): A \textbf{conservative resample} has too many large $p$-values and occurs when the bootstrap resamples are too similar to the observed, which would lead to underestimation of the variance. 
}
\label{figure:interprettingCurves}
\end{figure}

Now that we have introduced our bootstrap validation procedure, we state our main theorem.

\textbf{Theorem 1.} Let $\mA^{(1)}, \dots, \mA^{(M)}$ be a collection of BIRGs with probability matrices $\mP^{(1)}, \dots, \mP^{(M)}$, respectively. Let $\widetilde{\mA}^{(1)}, \dots, \widetilde{\mA}^{(M)}$ be a collection of truly resampled networks, that is, $\widetilde{\mA}^{(m)}_{ij} \overset{\text{ind}}{\sim} \text{Bernoulli}(\mP^{(m)}_{ij})$ for each $i,j \in \{1, \dots, n\}$ and $m \in \{1, \dots, M\}$. Then, the distribution of the $M$ $p$-values computed by the bootstrap exchangeability test (Algorithm \ref{alg:test}) on each pair $(\mA^{(m)}, \widetilde{\mA}^{(m)})$ is uniform on $[0, 1]$.

For proof see Appendix \ref{app:proof_theorem1}. 
In words, this theorem states that in the case where a bootstrap is truly exchangeable with the observed, our bootstrap validation procedure will provably deem the bootstrap valid. 
Theorem 1 provides a framework for verifying whether a proposed bootstrapping procedure generates exchangeable networks, offering a robust tool for validating the effectiveness of the bootstrapping procedure. 


When applying Theorem 1, we obtain $M$ $p$-values, call these $\hat{p}_1,...,\hat{p}_M$. 
We can sort these $p$-values such that $\hat{p}_{1^\prime} \le ... \le \hat{p}_{M^\prime}$. Under the null hypothesis, the $p$-values follow a Uniform[0,1] distribution, thus we can associate the $m$-th sorted $p$-value $\hat{p}_{m^\prime}$ with the $m$-th quantile given by $q_{m'} = \frac{m^\prime}{M+1}$ for $m \in \{1,...,M\}$ . By plotting each sorted $p$-value with its associated quantile we build a quantile-quantile (QQ)-plot. 
We construct a `Bootstrap Validity Score' 

\begin{equation}\label{eq:bootstrap_validity_score}
    S = \frac{1}{M}\sum\limits_{m^\prime=1}^M \left| \hat{p}_{m^\prime} - q_{m^\prime} \right|,
\end{equation}
where the closer the score is to 0, the closer to uniform the $p$-values are. A small Bootstrap Validity Score implies the bootstrap procedure used creates exchangeable bootstraps. 


Figure \ref{figure:interprettingCurves} illustrates the possible outcomes of the testing procedure for the sorted $p$-values. A \textit{valid} bootstrap procedure has a QQ-plot close to the diagonal and $S \approx 0$. 
If $\widetilde{\mA}$ is far from $\mA$, then this curve appears super-uniform; we deem such a bootstrap to be \textit{invalid}. Invalid bootstraps may be overdispersed (i.e. the bootstrap embeddings of the node would imply a variance larger than the true variance) or mean-biased (the bootstrapped mean is not centred on the true latent position). 
On the other hand, if $\widetilde{\mA}$ is too similar to $\mA$, then the curve appears sub-uniform; in this case, the hypothesis testing literature would deem the test \textit{conservative}. The embeddings being underdispersed can give a conservative test result. 
Both invalid and conservative bootstraps are distinguishable from the observed data and node variances would be misestimated. Therefore, we recommend that only valid bootstraps be trusted for such downstream applications.




\begin{algorithm}[ht]
\caption{Bootstrap Exchangeability Test}\label{alg:test}
\begin{algorithmic}[1]
\State \textbf{Input:}
\State Observed network $\mathbf{A} \in \{0,1\}^{n \times n}$
\State Bootstrap network $\widetilde{\mathbf{A}} \in \{0,1\}^{n \times n}$
\State Embedding dimension $d \leq n$
\State Number of permutations $R$
\State Test statistic $T_{r+1}$ from Eq. \ref{eq:t_q}.
\State \textbf{Compute:}
\State \parbox[t]{\dimexpr\linewidth-\algorithmicindent}{Compute the $d$-dimensional UASE of $\mA$ and $\widetilde{\mA}$, $\hat{\mY} = (\hat{\mY}^{(\text{obs})}; \widetilde{\mY}) \in \R^{2n \times d}$}
\State \parbox[t]{\dimexpr\linewidth-\algorithmicindent}{Compute observed test statistic \\ $T_1 = t_{obs}= t(\hat{\mY}) =t(\hat{\mY}^{(obs)}, \widetilde{\mY})$}
\For{$r = 1$ to $R$}
    \For{$i = 1$ to $n$}
        \State \parbox[t]{\dimexpr0.9\linewidth-\algorithmicindent}{Swap rows $i$ and $i+n$ of $\hat{\mathbf{Y}}$ with probability $\frac{1}{2}$ to obtain $\pi_r(\hat{\mathbf{Y}})$}
    \EndFor
    \State \parbox[t]{\dimexpr0.8\linewidth-\algorithmicindent}{Compute permuted test statistic $T_{r+1} = t(\pi_r(\hat{\mY}))$}
\EndFor
\State {Compute $p$-value} using Eq. \ref{eq:p_val}.
\State \textbf{Output:} $ \hat{p} $ 
\end{algorithmic}
\end{algorithm}

\subsection{Na\"ive Bootstrap}\label{section:XYTbootstrap}



Here, we consider a natural na\"ive bootstrap method. 
For the moment, let us consider a different network model, a Random Dot Product Graph (RDPG) \citep{young2007random,nickel2006random}. 
An RDPG is a latent position model, meaning it models each node as having a true position in $\R^{d}$ latent space, and provides a model-based rationale for spectral embedding. The RDPG assumes that $\mP = \mX\mX^T$, where $\mX \in \R^{n \times d}$ contains, as rows, the $d$-dimensional latent position vectors for each node. 
It is therefore reasonable to consider the ASE estimator $\hat{\mP}= \hat{\mX}\hat{\mX}^T$ as in Lemma \ref{lemma1}.

\begin{lemma}\label{lemma1}
   Let $\hat{\mX} \in \mathbb{R}^{n\times d}$ be an ASE of an observed adjacency matrix $\mA \in \mathbb{R}^{n \times n}$ and let $\mP \in [0, 1]^{n \times n}$ be the corresponding rank-$d$ probability matrix. 
   Assuming that $O(\|\hat{\mX}_i\|), O(\|\mX_i\|) = O(\mathrm{polylog}(n))$ for all $i$, 
   \begin{equation*}
       \underset{n \to \infty}{lim} ||\hat{\mX}\hat{\mX}^T - \mP|| = 0.
   \end{equation*}
\end{lemma}

For proof see Appendix \ref{proofoflemma1}. 
However, the estimator $\hat{\mP}= \hat{\mX}\hat{\mX}^T$ is problematic. Firstly, in the finite-sample case, the entries of $\hat{\mP} = \hat{\mX} \hat{\mX}^T$ can lie outside of the range $[0,1]$. 
If this occurs, we set any values less than 0 to 0 and any values greater than 1 to 1. Secondly, the assumptions are strong; for example, if $d$ needs to grow with $n$ then $\|\hat{X}_i\|$ grows as well. In practice \cite{levin2019bootstrapping} show this estimator performs badly in U-statistic bootstrap tests, as it is not asymptotically consistent. We will use it as a baseline to show that our test is capable of identifying problematic bootstraps. 



\subsection{Finite-sample kNN Bootstrap}\label{section:knnBootstrap}
We now propose a new procedure for creating network bootstraps,  
by estimating the behaviour of each node based on its local $k$-nearest neighbours (kNN). Using ASE, high-numbers of dimensions can be needed to correctly estimate neighbourhoods, which can make estimation hard. Since kNN relies solely on a locally linear structure that preserves nearest-neighbourhood relationships, alternative embeddings that reduce dimensionality can also be applied.

Specifically, we apply kNN to a single network embedding $\hat{\mX}$ to generate a $k$-length neighbourhood $\mathcal{N}_i \in \{1, \dots, n\}^k$ for each $i \in \{1, \dots, n\}$, where $i \in \mathcal{N}_i $ for all $i\in\{1,...,n\}$. For the examples in this paper, Euclidean distance 
is used as the distance measure. 
The $i$-th row of the estimated probability matrix $\mP \in \mathbb{R}^{n \times n}$ is given by the mean of the rows of $\mA$ indexed by the neighbours of node $i$,
\[\hat{\mP}_i = \frac{1}{k} \sum\limits_{j \in \mathcal{N}_i} \mA_j
\]
for each $i \in \{1, \dots, n\}$. 
Networks are then sampled from a BIRG with a probability matrix $\hat{\mP}$. 

\begin{algorithm}[ht]
\caption{ASE-kNN}\label{alg:knn_bootstrap}
\begin{algorithmic}[1]
\State \textbf{Input:}
\State Observed network $\mA \in \{0,1\}^{n \times n}$
\State Embedding dimension $d \leq n$
\State Number of nearest neighbours $k>1$
\State Number of bootstrapped graphs $B$
\State Distance measure for the data and embedding method, e.g. Euclidean
\State \textbf{Compute:}
\State Compute the $d$-dimensional ASE $\hat{\mX}$ of $\mA$
\For {$i = 1, \dots, n$}
\State Find the $k$ nearest neighbours of node $i$ in $\hat{\mX}$, denoted by the set $\mathcal{N}_i \in \{1, \dots, n\}^k$ (including node $i$), using specified distance measure
\State Set $\hat{\mP}_{i} = \frac{1}{k} \sum\limits_{j \in \mathcal{N}_i} \mA_{j}$
\EndFor
\For {$b=1,\dots, B$}
\State Sample  $\widetilde{\mA}^{(b)}_{ij} \overset{\text{ind}}{\sim} \text{Bernoulli}(\hat{\mP}_{ij})$\\
for each $i, j \in \{1, \dots, n\}$
\EndFor
\State \textbf{Output:} $\widetilde{\mA}^{(1)},\dots,\widetilde{\mA}^{(B)}$
\end{algorithmic}
\end{algorithm}

\begin{figure*}
    \centering
    \includegraphics[width=\linewidth]{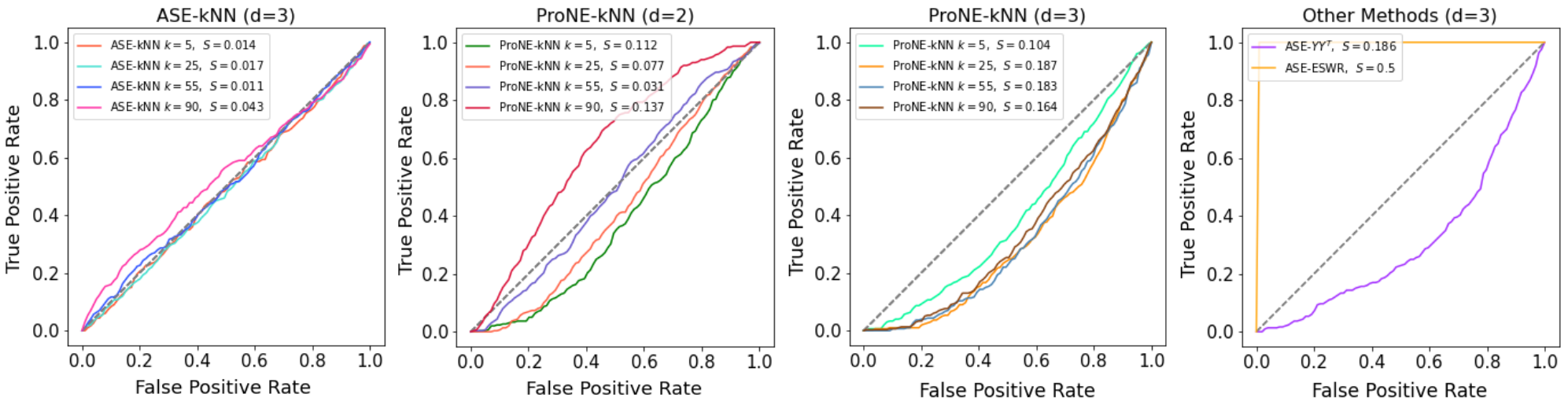}
    \caption{Simulated results for the 3 community MMSBM ($n=300$) with different embedding and bootstrap resampling methods applied. Each curve uses $M=300$ draws from the known model, each paired with a single bootstrap from the model. The area between the curve and the $x=y$ line is the Bootstrap Validity Score $S$.}
    \label{fig:MMSBMex}
\end{figure*}

For spectral embedding, an alternative form of $\hat{\mX}$, given in Appendix \ref{alt_Y_ASE}, can be used, which is less sensitive to the choice of $d$. 

%
When choosing $k$ there is a trade-off. 
If $k$ is too small we may be ignoring neighbours with similar behaviour, resulting in a bootstrap that is too similar to the observation. 
However, if $k$ is too large the model will use 
nodes embedded far away from the node of interest to estimate the latent position of the node of interest (see Figure \ref{figure:interprettingCurves}). 
Appendix \ref{appendix:knn_sensitivity_k} explores the sensitivity of this procedure to the choice of $k$. 
We check the appropriateness of the choice of $k$ by considering the Bootstrap Validity Score $S$.

\section{SYNTHETIC DATA EXAMPLE}\label{section:synthetic_data_example}

In this section we empirically evaluate our proposed network procedures alongside other bootstrapping procedures using the Bootstrap Exchangeability Test given in Section \ref{section:bootstrap_procedure_validation}. 
We use synthetic data generated from a Mixed Membership Stochastic Block Model (MMSBM) \citep{airoldi2008mixed}. 
MMSBMs are an extension of the stochastic block model (SBM) \citep{holland1983stochastic} that allow nodes to have partial membership to multiple communities. In an MMSBM, each node is associated with a distribution over communities, rather than being assigned to a single community. 
A vector $\bm{\alpha} \in \sR_+^C$ controls the mixed-membership allowed per node. 
For each node $i \in \{1,...,n\}$ there is a $C$-dimensional mixed membership vector $\bm{\pi}_i\overset{\text{ind}}{\sim}\text{Dirichlet}(\bm{\alpha})$.

An indicator vector $\vz_{i\rightarrow j}$ denotes the specific block membership of node $i$ when it connects to node $j$ and $\vz_{i\rightarrow j}$ denotes the specific block membership of node $j$ when it is connected from node $i$. 
For each pair of nodes $(i,j)$, a membership indicator for the initiator, $\vz_{i\rightarrow j} \overset{\text{ind}}{\sim} \text{Multinomial}(\bm{\pi}_i)$, is drawn, and a membership indicator for the receiver, $\vz_{j\rightarrow i} \overset{\text{ind}}{\sim} \text{Multinomial}(\bm{\pi}_j)$, is drawn. 
A block probability matrix $\mB \in [0,1]^{C \times C}$ defines the probability of interactions between the $C$ communities, where $\mB_{gh}$ is the probability of there being a connection from a node in community $g$ to a node in community $h$, for $g,h\in\{1,...,C\}$. 
By defining $\mB$ to be symmetric, $\mB_{gh}=\mB_{hg}$ for all $g,h\in\{1,...,C\}$, we have that $\mathbb{P}(\mA_{ij}) = \mathbb{P}(\mA_{ji})$ for all $i,j \in \{1,...,n\}$, however this does not guarantee that $\mA_{ij}=\mA_{ji}$. To ensure sampled matrices are symmetric, whenever $i>j$, 
we set $\mA_{ij} = \mA_{ji}$. 
The value of the interaction from node $i$ to node $j$ is sampled as 
\begin{equation}\label{eq:MMSBM}
    \mA_{ij} \overset{\text{ind}}{\sim} \text{Bernoulli}(\vz_{i\rightarrow j} \mB \vz_{j\rightarrow i}),
\end{equation} 
for all $i,j \in \{1,...,n\}$. 
In this example, there are $n=300$ nodes and $C=3$ communities. We define $\bm{\alpha}=\vone_C$, that is, each node is equally likely to belong to each community. 
We define the block probability matrix $\mB$ as: 
\begin{equation}\label{eq:B_for_MMSBM}
    \mB = 
    \begin{bmatrix}
    0.3 & 0.2 & 0.2 \\
    0.2 & 0.6 & 0.2 \\
    0.2 & 0.2 & 0.9 \\
    \end{bmatrix}.   
\end{equation}

Here we demonstrate the method is applicable to non clustered data; see Appendix \ref{app:4commSBMex} for a synthetic example on clustered data\footnote{We provide all data and implementation code for the synthetic and real-data experiments described in this paper at: \url{https://github.com/0emerald/ValidBootstrapsForNetworkEmbeddings}}. 

As introduced in Section \ref{section:bootstrap_procedure_validation}, we can use Algorithm \ref{alg:test} $M \in \mathbb{N}$ times to verify if a procedure for generating bootstrapped networks creates exchangeable networks. 
For this example we take $M=300$ random samples from the model specified by Equations \ref{eq:MMSBM} and \ref{eq:B_for_MMSBM} to create the adjacency matrices $\mA^{(1)},...,\mA^{(M)}$, and choose $d=3$ for ASE since rank$(\mB)=3$. 
For each matrix one bootstrap resample is generated as in Algorithm \ref{alg:knn_bootstrap} with $k=5$, given by $\widetilde{\mA}^{(1)},...,\widetilde{\mA}^{(M)}$, respectively. 
To each pair $[\mA^{(m)}, \widetilde{\mA}^{(m)}]$, for $m=1,...,M$, we apply Algorithm \ref{alg:test}, and use the $M$ $\hat{p}$-values to give a QQ-plot and a Bootstrap Validity Score. 
We repeat the kNN-based bootstrap with different values of $k$ to demonstrate the effect of $k$ in the model. Other bootstrap methods are also evaluated on this data (see Appendix \ref{appendix:used_bootstrap_methods}), in addition to using ProNE \citep{Zhang2019Prone} to create non-linear low-dimensional embeddings to apply bootstrap methods to. 

In this example, we use a variety of methods to estimate $\hat{\mP}$ from one network observation and draw bootstraps, which we then evaluate with the validity test (observed visually with the QQ-plots in Figure \ref{fig:MMSBMex}) to decide which method for estimating $\hat{\mP}$ gives exchangeable embeddings with the one observation. We are then free to use any embedding bootstraps that do pass the test for any downstream tasks. 

Figure \ref{fig:MMSBMex} shows that only our kNN methods achieve validity, though non-linear embeddings require more tuning as they compress information into fewer dimensions. Appendix \ref{app:MMSBM_n} considers the effects of changing $n$ in this model.

\section{REAL-WORLD DATA}\label{section:real_data}

\subsection{School Social Interaction Network}

For a concrete analysis on real data we consider a social interaction network. The Lyon school dataset captures face-to-face interactions between members of a French primary school over two days in October 2009, from 08:00 to 18:00 each day \citep{stehle2011high, rossi2015network}. The network tracks the interactions of 242 participants (10 teachers and 232 students) across 5 school years, each year group having two classes. Each person wore a Radio Frequency Identification (RFID) 
badge which recorded an interaction between two people if they were in close proximity for 20 seconds. 

\begin{figure}[!htb]
    \centering
    \includegraphics[width=\linewidth]{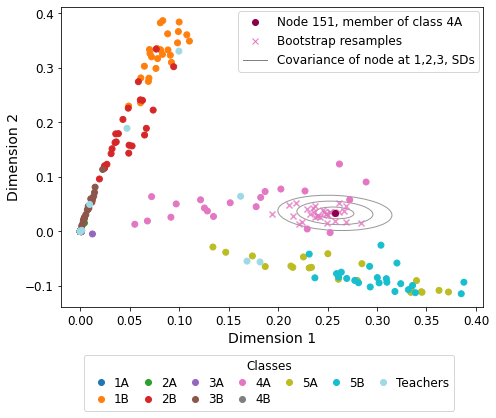}
    \caption{$\hat{\mY}^{(\text{obs})}$  the first 2 dimensions of the School data for all nodes. For a student node in class 4A, we visualise $B=20$ bootstrap embeddings, with the covariance ellipse at 1, 2, and 3 standard deviations. }
    \label{fig:school_A1_embedA_viaUASE_node151highlight}
\end{figure}

The data is binned into hour-long time windows over the two days to give a time series of networks. 
When two participants had one or more interactions in the hour-long window, an edge is present between them in the corresponding network. 
This gives a series of graphs $\mA^{(1)},...,\mA^{(20)} \in \{0,1\}^{n \times n}$, where $n=242$. 

To highlight the performance and a useful application of the methodology in this paper, we consider a single from the series, 9:00-10:00 of Day 1. This is $\mA^{(2)}$, which we shall henceforward denote as $\mA^{(obs)}$. 
We apply our bootstrap method to this data as in Algorithm \ref{alg:knn_bootstrap} to obtain $B=500$ bootstrap replicates of $\mA^{(obs)}$. 
We embed the observed matrix along with the $B=500$ bootstraps into a common $d=12$ dimensional space via UASE of $\bmgA = \left(\mA^{(obs)}, \widetilde{\mA}^{(1)},...,\widetilde{\mA}^{(B)}\right)$, to get an embedding for the observed matrix and the $B$ bootstrap matrices, denoted by $\hat{\mY}^{(\text{obs})}, \widetilde{\mY}^{(1)}, ..., \widetilde{\mY}^{(B)} \in \sR^{n \times d}$, respectively. 
From these, we are able to estimate the covariance of each node in the embedding space. 
Figure \ref{fig:school_A1_embedA_viaUASE_node151highlight} shows the first 2 dimensions of $\hat{\mY}^{(\text{obs})}$ with bootstrap embeddings and Normal-approximation uncertainty for one node.

\subsection{Applying Different Embedding Methods to the School Data
}

\begin{figure}[!htb]
    \centering
    \includegraphics[width=1.03\linewidth]{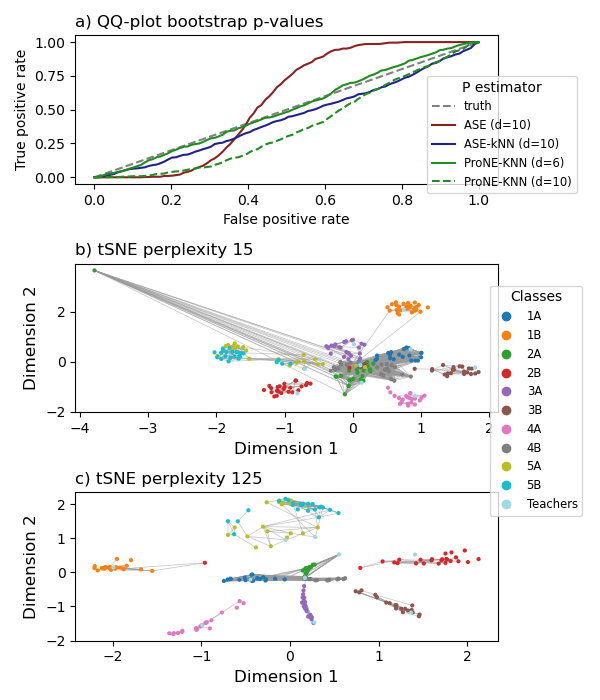}
    \caption{Insights from uncertainty in the School data. (a): QQ-plots from applying the Bootstrap Exchangeability Test to different bootstrap methods which use different estimators of $\hat{\mP}$. 
    (b): t-SNE of $\mA^{(obs)}$ using too-low perplexity results in distant points having overlapping uncertainty, shown as lines connecting nodes within 3 standard deviations of one another ($\hat{\mP}$ from $d=6$ ProNE embedding). 
    (c): t-SNE of $\mA^{(obs)}$ using higher perplexity results in fewer poorly placed nodes. } 
    \label{fig:tSNE_school}
\end{figure}

Figure \ref{fig:tSNE_school}a shows that the empirical performance of different estimators of $\hat{\mP}$ varies considerably. Despite asymptotic guarantees, ASE-based $\mX\mX^T$ approaches create bootstraps which the Bootstrap Exchangeability Test do not deem exchangeable. 
Bootstraps from nonlinear methods perform better; both kNN applied to the ASE embedding (ASE-kNN) and kNN applied to the 
ProNE embedding (ProNE-kNN). 

Our bootstrap procedure generates uncertainty for each point in the embedding, which we have providing a formal test to ensure validity. One important application is to score common network embeddings in terms of whether they represent this uncertainty. 
Dimension reduction techniques that preserve global structure in 2-3D such as t-SNE \citep{van2008visualizing} and UMAP \citep{mcinnes2018umap} are often applied. See Appendix \ref{tsneOverview} for a brief overview of t-SNE. 
We apply t-SNE to $\mA^{(obs)}$ to visualise the adjacency matrix in 2-dimensions. 
Bootstrapping $\mA^{(obs)}$ allows us to estimate the covariance of each node in the $d=10$ dimensional embedding $\hat{\mY}^{(obs)}$ which shows whether dimension reduction method such as t-SNE is appropriate. 
To do this we construct a symmetric ``fuzziness'' matrix $\mF \in \{0,1\}^{n \times n}$, where 
\[
\mF_{ij} = 
\begin{cases} 
\raisebox{2.3em}{1} & \text{\shortstack[l]{if node \( i \) is within 3 standard deviations \\[1pt]  
    of node \( j \) and node \( j \) is within 3 standard \\[1pt]  
    deviations of node \( i \),}} \\
0 & \text{otherwise.}
\end{cases}
\]
An edge is drawn between nodes $i$ and $j$ in the t-SNE plot if $\mF_{ij}=1$, with the interpretation that these nodes have overlapping probability distributions. Appendix \ref{robustnessChecks_tSNE_school} conducts sensitivity analysis to ensure robustness to t-SNE hyperparameters and the number of bootstraps $B$. 

To use network uncertainty to quantify the performance of some (externally chosen) embedding, we can then provide a score for a particular (standardised) node visualisation layout $\mV$, which rewards placing nodes with overlapping distributions close to one another, whilst also retaining structure within uncertain nodes by penalising nodes that are placed too close. For this we construct the `Fuzziness Score' as the standard deviation of the edge length 
\begin{equation}\label{eq:fuzzinessScore}
    F(\mV) = \mathrm{Var}(|\mV_i-\mV_j|_2 ; \mF_{ij}=1).
\end{equation}
This penalises both misplacing nodes in the wrong cluster, and giving uncertain nodes `structure'. 
To select a value for the perplexity hyperparameter we minimise the `Fuzziness score'. 
This implies an optimal perplexity hyperparameter for t-SNE on the School data (of 125$\pm$75; Figure \ref{fig:tsne_perplexity}).

Figure \ref{fig:tSNE_school}b-c shows $\mF$ as edges highlighting which nodes probabilistically overlap. Whilst the latent positions $\mY_i$ and $\mY_j$ may not be close, if $\mF_{ij}=1$ then it is possible that $\hat{\mY}_i$ and $\hat{\mY}_j$ could be the same. 
Using kNN bootstraps, we quantify the performance of t-SNE embeddings (Figure \ref{fig:tSNE_school}b-c) using node-wise variance. Annotating the fuzziness highlights across-cluster interactions that t-SNE alone does not show. 
The low perplexity visualisation is misleading - e.g. it places a class 2A student (green) away from students with overlapping distributions under bootstrap, and similarly a class 2B student (red) is placed centrally. Conversely, the high perplexity solution avoids this problem, whilst uncertainty highlights that the red 2B student's distribution overlaps with 1B.


\section{DISCUSSION}\label{section:discussion_conclusion}

Obtaining valid bootstrap resamples from a network embedding requires only a single substantial assumption, that nodes are exchangeable. 
This permits versatile resampling from a single network observation, validated by a reliable Bootstrap Exchangeability Test to confirm resample exchangeability with the observed network. 
Our approach relies on $\mP$ being a sufficient statistic for $\mA^{(obs)}$, so extending the methodology to work for weighted networks would require additional distributional assumptions. 
Experimental evaluation demonstrates the method is able to resample networks of various structures. 
The bootstrap validity test supports the flexibility to incorporate various embedding methods, including those optimised for scalability with large datasets, ensuring applicability to networks with increasing complexity and size. 

In the common case where only one instance of a network is available, care must be taken not to overfit. The fuzziness score and other metrics should be viewed as a sensitivity analysis, and not subject to premature optimisation. The number of neighbours $k$ should be chosen by a prior based on community sizes, and run with other $k$ used to ensure a lack of sensitivity to this parameter. 

Previous studies found that using the estimate $\hat{\mP} = \hat{\mX}\hat{\mX}^T$ introduces bias in count statistics. Our findings extend this by showing it does not provide a valid bootstrap under the exchangeable bootstrap framework proposed in this paper. 
In contrast, the ASE-kNN bootstrap we introduce generates exchangeable embeddings, but it underestimates the variability in metrics such as average node degree and triangle density in an SBM example. 

Future work aimed at developing network bootstrapping methods that satisfy all validity criteria could enable broader applications in downstream tasks. By offering a novel perspective on evaluating bootstrap validity, we address a range of network analysis applications that rely on low-dimensional representations rather than entire networks. 




\bibliography{references}

\begin{thebibliography}{53}
\providecommand{\natexlab}[1]{#1}
\providecommand{\url}[1]{\texttt{#1}}
\expandafter\ifx\csname urlstyle\endcsname\relax
  \providecommand{\doi}[1]{doi: #1}\else
  \providecommand{\doi}{doi: \begingroup \urlstyle{rm}\Url}\fi

\bibitem[Airoldi et~al.(2008)Airoldi, Blei, Fienberg, and Xing]{airoldi2008mixed}
Edo~M Airoldi, David Blei, Stephen Fienberg, and Eric Xing.
\newblock Mixed membership stochastic blockmodels.
\newblock \emph{Advances in neural information processing systems}, 21, 2008.

\bibitem[Akoglu et~al.(2015)Akoglu, Tong, and Koutra]{akoglu2015graph}
Leman Akoglu, Hanghang Tong, and Danai Koutra.
\newblock Graph based anomaly detection and description: a survey.
\newblock \emph{Data mining and knowledge discovery}, 29:\penalty0 626--688, 2015.
\newblock \doi{10.1007/s10618-014-0365-y}.

\bibitem[Bhattacharyya and Bickel(2015)]{bhattacharyya2015subsampling}
Sharmodeep Bhattacharyya and Peter~J. Bickel.
\newblock {Subsampling bootstrap of count features of networks}.
\newblock \emph{The Annals of Statistics}, 43\penalty0 (6):\penalty0 2384 -- 2411, 2015.
\newblock \doi{10.1214/15-AOS1338}.

\bibitem[Bilot et~al.(2023)Bilot, El~Madhoun, Al~Agha, and Zouaoui]{bilot2023graph}
Tristan Bilot, Nour El~Madhoun, Khaldoun Al~Agha, and Anis Zouaoui.
\newblock Graph neural networks for intrusion detection: A survey.
\newblock \emph{IEEE Access}, 2023.
\newblock \doi{10.1109/ACCESS.2023.3275789}.

\bibitem[Borgs et~al.(2008)Borgs, Chayes, Lov{\'a}sz, S{\'o}s, and Vesztergombi]{borgs2008convergent}
Christian Borgs, Jennifer~T Chayes, L{\'a}szl{\'o} Lov{\'a}sz, Vera~T S{\'o}s, and Katalin Vesztergombi.
\newblock Convergent sequences of dense graphs i: Subgraph frequencies, metric properties and testing.
\newblock \emph{Advances in Mathematics}, 219\penalty0 (6):\penalty0 1801--1851, 2008.
\newblock \doi{10.1016/j.aim.2008.07.008}.

\bibitem[Bowman and Huang(2021)]{bowman2021towards}
Benjamin Bowman and H~Howie Huang.
\newblock Towards next-generation cybersecurity with graph ai.
\newblock \emph{ACM SIGOPS Operating Systems Review}, 55\penalty0 (1):\penalty0 61--67, 2021.
\newblock \doi{10.1145/3469379.3469386}.

\bibitem[Cape et~al.(2019)Cape, Tang, and Priebe]{Cape2019TwoToInfinity}
Joshua Cape, Minh Tang, and Carey~E. Priebe.
\newblock The two-to-infinity norm and singular subspace geometry with applications to high-dimensional statistics.
\newblock \emph{The Annals of Statistics}, 47\penalty0 (5):\penalty0 2405--2439, 2019.
\newblock ISSN 00905364, 21688966.
\newblock \doi{10.48550/arXiv.1705.10735}.

\bibitem[Chen et~al.(2020)Chen, Arroyo, Athreya, Cape, Vogelstein, Park, White, Larson, Yang, and Priebe]{chen2020multiple}
Guodong Chen, Jes{\'u}s Arroyo, Avanti Athreya, Joshua Cape, Joshua~T Vogelstein, Youngser Park, Chris White, Jonathan Larson, Weiwei Yang, and Carey~E Priebe.
\newblock Multiple network embedding for anomaly detection in time series of graphs.
\newblock \emph{arXiv preprint arXiv:2008.10055}, 2020.
\newblock \doi{10.48550/arXiv.2008.10055}.

\bibitem[Davis et~al.(2023)Davis, Gallagher, Lawson, and Rubin-Delanchy]{davis2023simple}
Ed~Davis, Ian Gallagher, Daniel~John Lawson, and Patrick Rubin-Delanchy.
\newblock A simple and powerful framework for stable dynamic network embedding.
\newblock \emph{arXiv preprint arXiv:2311.09251}, 2023.
\newblock \doi{10.48550/arXiv.2311.09251}.

\bibitem[Davis et~al.(2024)Davis, Gallagher, Lawson, and Rubin-Delanchy]{davis2024valid}
Ed~Davis, Ian Gallagher, Daniel~John Lawson, and Patrick Rubin-Delanchy.
\newblock Valid conformal prediction for dynamic gnns.
\newblock \emph{arXiv preprint arXiv:2405.19230}, 2024.
\newblock \doi{10.48550/arXiv.2405.19230}.

\bibitem[De~Lathauwer et~al.(2000)De~Lathauwer, De~Moor, and Vandewalle]{Lathauwer2000SVD}
Lieven De~Lathauwer, Bart De~Moor, and Joos Vandewalle.
\newblock A multilinear singular value decomposition.
\newblock \emph{SIAM Journal on Matrix Analysis and Applications}, 21\penalty0 (4):\penalty0 1253--1278, 2000.
\newblock \doi{10.1137/S0895479896305696}.

\bibitem[Efron(1979)]{efron1979bootstrap}
B.~Efron.
\newblock {Bootstrap Methods: Another Look at the Jackknife}.
\newblock \emph{The Annals of Statistics}, 7\penalty0 (1):\penalty0 1 -- 26, 1979.
\newblock \doi{10.1214/aos/1176344552}.

\bibitem[Efron and Tibshirani(1994)]{efron1994introduction}
Bradley Efron and Robert~J Tibshirani.
\newblock \emph{An introduction to the bootstrap}.
\newblock Chapman and Hall/CRC, 1994.
\newblock \doi{10.1201/9780429246593}.

\bibitem[Epskamp et~al.(2018)Epskamp, Borsboom, and Fried]{epskamp2018estimating}
Sacha Epskamp, Denny Borsboom, and Eiko~I Fried.
\newblock Estimating psychological networks and their accuracy: A tutorial paper.
\newblock \emph{Behavior research methods}, 50:\penalty0 195--212, 2018.
\newblock \doi{10.3758/s13428-017-0862-1}.

\bibitem[Gallagher et~al.(2021)Gallagher, Jones, and Rubin-Delanchy]{gallagher2021spectral}
Ian Gallagher, Andrew Jones, and Patrick Rubin-Delanchy.
\newblock Spectral embedding for dynamic networks with stability guarantees.
\newblock \emph{Advances in Neural Information Processing Systems}, 34:\penalty0 10158--10170, 2021.
\newblock \doi{10.48550/arXiv.2106.01282}.

\bibitem[Gao et~al.(2022)Gao, Yu, and Zhang]{gao2022clustering}
Yang Gao, Xiangzhan Yu, and Hongli Zhang.
\newblock Graph clustering using triangle-aware measures in large networks.
\newblock \emph{Information Sciences}, 584:\penalty0 618--632, 2022.
\newblock ISSN 0020-0255.
\newblock \doi{10.1016/j.ins.2021.11.008}.

\bibitem[Green and Shalizi(2022)]{green2022bootstrapping}
Alden Green and Cosma~Rohilla Shalizi.
\newblock Bootstrapping exchangeable random graphs.
\newblock \emph{Electronic Journal of Statistics}, 16\penalty0 (1):\penalty0 1058--1095, 2022.
\newblock \doi{10.1214/21-EJS1896}.

\bibitem[Grover and Leskovec(2016)]{grover2016node2vec}
Aditya Grover and Jure Leskovec.
\newblock node2vec: Scalable feature learning for networks.
\newblock In \emph{Proceedings of the 22nd ACM SIGKDD international conference on Knowledge discovery and data mining}, pages 855--864, 2016.
\newblock \doi{10.48550/arXiv.1607.00653}.

\bibitem[He et~al.(2022)He, Ji, and Huang]{he2022illuminati}
Haoyu He, Yuede Ji, and H~Howie Huang.
\newblock Illuminati: Towards explaining graph neural networks for cybersecurity analysis.
\newblock In \emph{2022 IEEE 7th European Symposium on Security and Privacy (EuroS\&P)}, pages 74--89. IEEE, 2022.
\newblock \doi{10.1109/EuroSP53844.2022.00013}.

\bibitem[Hoff et~al.(2002)Hoff, Raftery, and Handcock]{hoff2002latent}
Peter~D Hoff, Adrian~E Raftery, and Mark~S Handcock.
\newblock Latent space approaches to social network analysis.
\newblock \emph{Journal of the american Statistical association}, 97\penalty0 (460):\penalty0 1090--1098, 2002.
\newblock \doi{10.1198/016214502388618906}.

\bibitem[Holland et~al.(1983)Holland, Laskey, and Leinhardt]{holland1983stochastic}
Paul~W Holland, Kathryn~Blackmond Laskey, and Samuel Leinhardt.
\newblock Stochastic blockmodels: First steps.
\newblock \emph{Social networks}, 5\penalty0 (2):\penalty0 109--137, 1983.
\newblock \doi{10.1016/0378-8733(83)90021-7}.

\bibitem[Jones and Rubin-Delanchy(2020)]{jones2020multilayerRDPG}
Andrew Jones and Patrick Rubin-Delanchy.
\newblock The multilayer random dot product graph.
\newblock \emph{arXiv}, page arXiv:2007.10455, 2020.
\newblock \doi{10.48550/ARXIV.2007.10455}.

\bibitem[Jumper et~al.(2021)Jumper, Evans, Pritzel, Green, Figurnov, Ronneberger, Tunyasuvunakool, Bates, {\v{Z}}{\'\i}dek, Potapenko, et~al.]{jumper2021highly}
John Jumper, Richard Evans, Alexander Pritzel, Tim Green, Michael Figurnov, Olaf Ronneberger, Kathryn Tunyasuvunakool, Russ Bates, Augustin {\v{Z}}{\'\i}dek, Anna Potapenko, et~al.
\newblock Highly accurate protein structure prediction with alphafold.
\newblock \emph{Nature}, 596\penalty0 (7873):\penalty0 583--589, 2021.
\newblock \doi{10.1038/s41586-021-03819-2}.

\bibitem[Koutra et~al.(2013)Koutra, Vogelstein, and Faloutsos]{koutra2013deltacon}
Danai Koutra, Joshua~T Vogelstein, and Christos Faloutsos.
\newblock Deltacon: A principled massive-graph similarity function.
\newblock In \emph{Proceedings of the 2013 SIAM international conference on data mining}, pages 162--170. SIAM, 2013.
\newblock \doi{10.1137/1.9781611972832.18}.

\bibitem[Kreiss and Paparoditis(2011)]{kreiss2011bootstrap}
Jens-Peter Kreiss and Efstathios Paparoditis.
\newblock Bootstrap methods for dependent data: A review.
\newblock \emph{Journal of the Korean Statistical Society}, 40\penalty0 (4):\penalty0 357--378, 2011.
\newblock \doi{10.1016/j.jkss.2011.08.009}.

\bibitem[Kullback and Leibler(1951)]{kullback1951information}
Solomon Kullback and Richard~A Leibler.
\newblock On information and sufficiency.
\newblock \emph{The annals of mathematical statistics}, 22\penalty0 (1):\penalty0 79--86, 1951.

\bibitem[Levin and Levina(2019)]{levin2019bootstrapping}
Keith Levin and Elizaveta Levina.
\newblock Bootstrapping networks with latent space structure.
\newblock \emph{arXiv preprint arXiv:1907.10821}, 2019.
\newblock \doi{10.48550/arXiv.1907.10821}.

\bibitem[Levin et~al.(2017)Levin, Athreya, Tang, Lyzinski, and Priebe]{levin2017central}
Keith Levin, Avanti Athreya, Minh Tang, Vince Lyzinski, and Carey~E Priebe.
\newblock A central limit theorem for an omnibus embedding of multiple random dot product graphs.
\newblock In \emph{2017 IEEE international conference on data mining workshops (ICDMW)}, pages 964--967. IEEE, 2017.
\newblock \doi{10.48550/arXiv.1705.09355}.

\bibitem[Lian(2011)]{lianConvergenceFunctionalKnearest2011}
Heng Lian.
\newblock Convergence of functional k-nearest neighbor regression estimate with functional responses.
\newblock \emph{Electronic Journal of Statistics}, 5:\penalty0 31--40, 2011.
\newblock ISSN 1935-7524, 1935-7524.
\newblock \doi{10.1214/11-EJS595}.

\bibitem[Lin et~al.(2008)Lin, Chi, Zhu, Sundaram, and Tseng]{lin2008facetnet}
Yu-Ru Lin, Yun Chi, Shenghuo Zhu, Hari Sundaram, and Belle~L Tseng.
\newblock Facetnet: a framework for analyzing communities and their evolutions in dynamic networks.
\newblock In \emph{Proceedings of the 17th international conference on World Wide Web}, pages 685--694, 2008.
\newblock \doi{10.1145/1367497.1367590}.

\bibitem[Lov{\'a}sz and Szegedy(2006)]{lovasz2006limits}
L{\'a}szl{\'o} Lov{\'a}sz and Bal{\'a}zs Szegedy.
\newblock Limits of dense graph sequences.
\newblock \emph{Journal of Combinatorial Theory, Series B}, 96\penalty0 (6):\penalty0 933--957, 2006.
\newblock \doi{10.1016/j.jctb.2006.05.002}.

\bibitem[McInnes et~al.(2018)McInnes, Healy, and Melville]{mcinnes2018umap}
Leland McInnes, John Healy, and James Melville.
\newblock Umap: Uniform manifold approximation and projection for dimension reduction.
\newblock \emph{arXiv preprint arXiv:1802.03426}, 2018.
\newblock \doi{10.48550/arXiv.1802.03426}.

\bibitem[Mikolov et~al.(2013)Mikolov, Chen, Corrado, and Dean]{word2vec}
Tomas Mikolov, Kai Chen, Greg Corrado, and Jeffrey Dean.
\newblock Efficient estimation of word representations in vector space.
\newblock \emph{arXiv preprint arXiv:1301.3781}, 2013.
\newblock \doi{10.48550/arXiv.1301.3781}.

\bibitem[Nguyen and Holmes(2019)]{screePlotElbow}
Lan Nguyen and Susan Holmes.
\newblock Ten quick tips for effective dimensionality reduction.
\newblock \emph{PLOS Computational Biology}, 15:\penalty0 e1006907, 06 2019.
\newblock \doi{10.1371/journal.pcbi.1006907}.

\bibitem[Nickel(2006)]{nickel2006random}
Christine Leigh~Myers Nickel.
\newblock \emph{Random dot product graphs a model for social networks}.
\newblock PhD thesis, Johns Hopkins University, 2006.

\bibitem[Pearson(1901)]{pearson1901PCA}
Karl Pearson.
\newblock Liii. on lines and planes of closest fit to systems of points in space.
\newblock \emph{The London, Edinburgh, and Dublin Philosophical Magazine and Journal of Science}, 2\penalty0 (11):\penalty0 559--572, 1901.
\newblock \doi{10.1080/14786440109462720}.

\bibitem[Prat-P\'{e}rez et~al.(2014)Prat-P\'{e}rez, Dominguez-Sal, and Larriba-Pey]{perez2014community}
Arnau Prat-P\'{e}rez, David Dominguez-Sal, and Josep-Lluis Larriba-Pey.
\newblock High quality, scalable and parallel community detection for large real graphs.
\newblock In \emph{Proceedings of the 23rd International Conference on World Wide Web}, WWW '14, page 225–236, New York, NY, USA, 2014. Association for Computing Machinery.
\newblock ISBN 9781450327442.
\newblock \doi{10.1145/2566486.2568010}.

\bibitem[Qin and Rohe(2013)]{qin2013regularized}
Tai Qin and Karl Rohe.
\newblock Regularized spectral clustering under the degree-corrected stochastic blockmodel.
\newblock \emph{Advances in neural information processing systems}, 26, 2013.
\newblock \doi{10.48550/arXiv.1309.4111}.

\bibitem[Rosenthal et~al.(2018)Rosenthal, V{\'a}{\v{s}}a, Griffa, Hagmann, Amico, Go{\~n}i, Avidan, and Sporns]{brain_embedding_node2vec}
Gideon Rosenthal, Franti{\v{s}}ek V{\'a}{\v{s}}a, Alessandra Griffa, Patric Hagmann, Enrico Amico, Joaqu{\'\i}n Go{\~n}i, Galia Avidan, and Olaf Sporns.
\newblock Mapping higher-order relations between brain structure and function with embedded vector representations of connectomes.
\newblock \emph{Nature communications}, 9\penalty0 (1):\penalty0 2178, 2018.
\newblock \doi{10.1038/s41467-018-04614-w}.

\bibitem[Rossi and Ahmed(2015)]{rossi2015network}
Ryan Rossi and Nesreen Ahmed.
\newblock The network data repository with interactive graph analytics and visualization.
\newblock In \emph{Proceedings of the AAAI conference on artificial intelligence}, volume~29, 2015.
\newblock \doi{10.1609/aaai.v29i1.9277}.

\bibitem[Rubin-Delanchy et~al.(2022)Rubin-Delanchy, Cape, Tang, and Priebe]{RubinDelanchy2022statistical}
Patrick Rubin-Delanchy, Joshua Cape, Minh Tang, and Carey~E. Priebe.
\newblock {A Statistical Interpretation of Spectral Embedding: The Generalised Random Dot Product Graph}.
\newblock \emph{Journal of the Royal Statistical Society Series B: Statistical Methodology}, 84\penalty0 (4):\penalty0 1446--1473, 06 2022.
\newblock ISSN 1369-7412.
\newblock \doi{10.1111/rssb.12509}.

\bibitem[Scheinerman and Tucker(2010)]{scheinerman2010modeling}
Edward~R Scheinerman and Kimberly Tucker.
\newblock Modeling graphs using dot product representations.
\newblock \emph{Computational statistics}, 25:\penalty0 1--16, 2010.
\newblock \doi{10.1007/s00180-009-0158-8}.

\bibitem[Seshadhri et~al.(2020)Seshadhri, Sharma, Stolman, and Goel]{seshadhri2020impossibility}
Comandur Seshadhri, Aneesh Sharma, Andrew Stolman, and Ashish Goel.
\newblock The impossibility of low-rank representations for triangle-rich complex networks.
\newblock \emph{Proceedings of the National Academy of Sciences}, 117\penalty0 (11):\penalty0 5631--5637, 2020.

\bibitem[S{\"o}derberg(2002)]{soderberg2002general}
Bo~S{\"o}derberg.
\newblock General formalism for inhomogeneous random graphs.
\newblock \emph{Physical review E}, 66\penalty0 (6):\penalty0 066121, 2002.
\newblock \doi{10.1103/PhysRevE.66.066121}.

\bibitem[Stehl{\'e} et~al.(2011)Stehl{\'e}, Voirin, Barrat, Cattuto, Isella, Pinton, Quaggiotto, Van~den Broeck, R{\'e}gis, Lina, et~al.]{stehle2011high}
Juliette Stehl{\'e}, Nicolas Voirin, Alain Barrat, Ciro Cattuto, Lorenzo Isella, Jean-Fran{\c{c}}ois Pinton, Marco Quaggiotto, Wouter Van~den Broeck, Corinne R{\'e}gis, Bruno Lina, et~al.
\newblock High-resolution measurements of face-to-face contact patterns in a primary school.
\newblock \emph{PloS one}, 6\penalty0 (8):\penalty0 e23176, 2011.
\newblock \doi{10.1371/journal.pone.0023176}.

\bibitem[Stone(1977)]{stoneConsistentNonparametricRegression1977}
Charles~J. Stone.
\newblock Consistent {{Nonparametric Regression}}.
\newblock \emph{The Annals of Statistics}, 5\penalty0 (4):\penalty0 595--620, 1977.
\newblock ISSN 0090-5364, 2168-8966.
\newblock \doi{10.1214/aos/1176343886}.

\bibitem[Sussman et~al.(2012)Sussman, Tang, Fishkind, and Priebe]{sussman2012consistentadjacencyspectralembedding}
Daniel~L. Sussman, Minh Tang, Donniell~E. Fishkind, and Carey~E. Priebe.
\newblock A consistent adjacency spectral embedding for stochastic blockmodel graphs, 2012.

\bibitem[Van~der Maaten and Hinton(2008)]{van2008visualizing}
Laurens Van~der Maaten and Geoffrey Hinton.
\newblock Visualizing data using t-sne.
\newblock \emph{Journal of machine learning research}, 9\penalty0 (11), 2008.

\bibitem[Wu et~al.(2022)Wu, Sun, Zhang, Xie, and Cui]{wu2022graph}
Shiwen Wu, Fei Sun, Wentao Zhang, Xu~Xie, and Bin Cui.
\newblock Graph neural networks in recommender systems: a survey.
\newblock \emph{ACM Computing Surveys}, 55\penalty0 (5):\penalty0 1--37, 2022.
\newblock \doi{10.1145/3535101}.

\bibitem[Young and Scheinerman(2007)]{young2007random}
Stephen~J Young and Edward~R Scheinerman.
\newblock Random dot product graph models for social networks.
\newblock In \emph{International Workshop on Algorithms and Models for the Web-Graph}, pages 138--149. Springer, 2007.
\newblock \doi{10.1007/978-3-540-77004-6_11}.

\bibitem[Zhang et~al.(2019)Zhang, Dong, Wang, Tang, and Ding]{Zhang2019Prone}
Jie Zhang, Yuxiao Dong, Yan Wang, Jie Tang, and Ming Ding.
\newblock Prone: Fast and scalable network representation learning.
\newblock In \emph{Proceedings of the Twenty-Eighth International Joint Conference on Artificial Intelligence, {IJCAI-19}}, pages 4278--4284. International Joint Conferences on Artificial Intelligence Organization, 7 2019.
\newblock \doi{10.24963/ijcai.2019/594}.

\bibitem[Zhu and Ghodsi(2006)]{ZHU2006918}
Mu~Zhu and Ali Ghodsi.
\newblock Automatic dimensionality selection from the scree plot via the use of profile likelihood.
\newblock \emph{Computational Statistics \& Data Analysis}, 51\penalty0 (2):\penalty0 918--930, 2006.
\newblock ISSN 0167-9473.
\newblock \doi{10.1016/j.csda.2005.09.010}.

\bibitem[Zu and Qin(2024)]{zu2024local}
Tianhai Zu and Yichen Qin.
\newblock Local bootstrap for network data.
\newblock \emph{Biometrika}, 112\penalty0 (1):\penalty0 asae046, 09 2024.
\newblock ISSN 1464-3510.
\newblock \doi{10.1093/biomet/asae046}.

\end{thebibliography}

\newpage
\newpage
\onecolumn

\clearpage

\newpage
\appendix
\onecolumn


\maketitle

\section{Appendix}

\subsection{Other Across-network Exchangeable Embeddings}\label{dilatedEmbeddingNetTest}

In Section \ref{section:UASE} we introduced UASE, which is a multi-network embedding that has the property of across-network exchangeability. This property is utilised in our bootstrap exchangeability test (stated in Algorithm \ref{alg:test}), in order to prove Theorem 1. 
UASE is just one of many possible \textit{unfolded} embeddings \citep{davis2023simple, davis2024valid}, which have across-network exchangeability.

The standard unfolding, used by UASE, is defined as a column concatenation of the collection of networks $\bmgA=(\mA^{(1)}, ..., \mA^{(M)}) \in \sR^{Mn \times n}$. However, it is possible to define a more general \textit{dilated unfolding}, 
\begin{equation}\label{eq:symmDilation_DUE}
    \mathscr{D}(\bmgA) = 
    \begin{bmatrix}
        \boldsymbol{0} & \bmgA \\
        \bmgA^T & \boldsymbol{0}
    \end{bmatrix}
    \in \{0,1\}^{n(M+1) \times n(M+1)}.
\end{equation}
Let $\mathcal{G} : \{\mA \in \{0, 1\}^{n \times n} : \mA = \mA^T\} \times \Omega \rightarrow \R^{n \times d}$ be a general function for single-network embedding of an adjacency matrix $\mA$ with a seed $\omega \in \Omega$. We require this seed due to the random nature of most network embedding methods; for example, the specific choice of eigenvalues and eigenvectors in ASE is random. The inclusion of some random seed serves the purpose of eliminating this randomness, allowing for comparisons to be made across multiple runs of the embedding function. This is required for the following definition. $\mathcal{G}$ is a label-invariant embedding in the case where 
\begin{equation*}
    \mathbb{P} (\mathcal{G}(\mathbf{a}, \omega) = \mathbf{v}) = \mathbb{P} \left( \mathcal{G}(\mathbf{\Pi a \Pi}^\top, \omega) = \mathbf{\Pi v}\right),
\end{equation*}
for any permutation matrix $\Pi \in \{0, 1\}^{n \times n}$, $\mathbf{a} \in \R^{n \times n}$ and $\mathbf{v} \in \R^{n \times d}$. \cite{davis2023simple} proves that any label-invariant $\mathcal{G}$, applied to $\mathscr{D}(\bmgA)$, returns 
\begin{equation*}
    \begin{bmatrix} \hat{\mX} \\ \hat{\mY} \end{bmatrix}
    = \mathcal{G} \left(\mathscr{D}(\bmgA), \omega \right),
\end{equation*}
where $\hat{\mX} \in \R^{n\times d}$ is a single embedding that summarises the collection of networks, and $\hat{\mY} \in \R^{Mn \times d}$ is an across-network exchangeable multi-network embedding. The authors show that when $\mathcal{G}$ is simply ASE, then the returned $\hat{\mY}$ is equivalent to UASE.

Using this more general definition of an unfolded embedding, we now have access to many other across-network exchangeable embeddings through the choice of $\mathcal{G}$. For example, we define unfolded ProNE, by setting $\mathcal{G}$ to be the ProNE embedding \citep{Zhang2019Prone}.

\subsection{Proof of Theorem 1}\label{app:proof_theorem1}
\begin{proof}
UASE is a multi-network embedding that satisfies across-network exchangeability (Definition 3) regardless of embedding dimension (\cite{gallagher2021spectral}, \cite{davis2023simple} Theorem 5). Let $(\hat{\mY}^{(m)}; \widetilde{\mY}^{(m)}) \in \R^{2n\times d}$ denote a $d$-dimensional UASE of $\bmgA = (\mA^{(m)}, \widetilde{\mA}^{(m)}) \in \mathbb{R}^{2n \times n}$ for all $m \in \{1,...,M\}$. 
Both $\mA^{(m)}$ and $\widetilde{\mA}^{(m)}$ are BIRGs drawn the same probability matrix, $\mP^{(m)}$. 
Since UASE is across-network exchangeable, it is true that 
\begin{align*}
    \mathbb{P} \left(\hat{\mY}^{(m)}_i = v_1, \widetilde{\mY}^{(m)}_i = v_2 \right) = 
    \mathbb{P} \left(\hat{\mY}^{(m)}_i = v_2, \widetilde{\mY}^{(m)}_i = v_1 \right).
\end{align*}
for all $m \in \{1,...,M\}$. In words, the embedding of the observed network $\mA^{(m)}$ is exchangeable with the embedding of the truly resampled network $\widetilde{\mA}^{(m)}$ for each $m \in \{1, \dots, M\}$.

Due to the exchangeability of each $\hat{\mY}^{(m)}, \widetilde{\mY}^{(m)}$, the permutations applied at each step $r \in \{1,...,R\}$ does not alter the distribution of $t(\hat{\mY}^{(m)}, \widetilde{\mY}^{(m)})$. Therefore, the sequence $T_1, T_2 \dots, T_{R+1}$ is exchangeable, that is 
\begin{equation*}
    \mathbb{P} (T_1 \leq v_1, \dots, T_{R+1} \leq v_{R+1} )= 
    \mathbb{P} (T_1 \leq v_{\sigma(1)}, \dots, T_{R+1} \leq v_{\sigma(R+1)})
\end{equation*}
for any $v_k \in \R$ and any permutation $\sigma$ on $\{1, \dots, R+1\}$.

As the sequence of test statistics is exchangeable, it follows that the $p$-value,
\begin{equation*}
    \hat{p} = \frac{1}{R+1} \sum\limits_{r=1}^{R+1} \mathbbm{1}\{T_r \ge t_{obs}\},
\end{equation*}
will be uniformly distributed on $[0,1]$. 
\end{proof}

\subsection{Proof of Lemma 1}\label{proofoflemma1}

\begin{proof}
   From \cite{gallagher2021spectral} (Proposition 2) we have that `there exists a sequence of orthogonal matrices $\mQ \in O(d)$ such that
   \begin{equation}\label{eq:6}
       \underset{i=1,...,n}{max} ||\hat{\mX}_i \mQ - \mX_i|| = O \left( \frac{\sqrt{\mathrm{log}(n)}}{\sqrt{\rho_n n}}\right)
   \end{equation}
   with high probability, where it is assumed that the sparsity factor $\rho_n$ satisfies $\rho_n = \omega \left( \frac{1}{n} \mathrm{log}^c(n) \right)$ for some universal constant $c > 1$'. 
   This states that after applying an orthogonal transformation to an embedding $\hat{\mX}$, which leaves the structure of the embedding unchanged, the embedding will converge in the Euclidean norm to the noise-free embedding $\mX$ as the number of nodes increases. For a dense graph $\rho_n =1$. 
   To prove the lemma, it is enough to assume that $O(\|\hat{\mX}_i\|), O(\|\mX_i\|) = O(\mathrm{polylog}(n))$ for all $i$. Under this assumption and using Equation \ref{eq:6} we obtain: 
   \begin{align}\label{eq:7}\nonumber
        \underset{i=1,...,n}{max}  \|\hat{\mX}_i \hat{\mX}_i^T - \mX_i \mX^T_i\| &\leq \underset{i=1,...,n}{max} \left( \|\hat{\mX}_i - \mX_i\| \cdot \|\hat{\mX}_i^T\| + \|\mX_i\| \cdot \|\hat{\mX}_i - \mX_i\| \right) \\ \nonumber
        &= O \left( \frac{\sqrt{\mathrm{log}(n)}}{\sqrt{\rho_n n}}\right) \cdot \underset{i=1,...,n}{max} \left( \|\hat{\mX}_i\| + \|\mX_i\| \right) \\ \nonumber
        &=  O \left( \frac{\sqrt{\mathrm{log}(n)} \cdot \mathrm{polylog}(n)}{\sqrt{\rho_n n}}\right) \\
   \end{align}

  The limit $\underset{n \to \infty}{lim} \frac{\sqrt{\mathrm{log}(n)}\cdot \mathrm{polylog}(n)}{\sqrt{\rho_n n}} = 0$, from which we get $\underset{n \to \infty}{lim} \underset{i=1,...,n}{max} ||\hat{\mX}_i \hat{\mX}_i^T - \mX_i \mX^{T}_i|| = 0$ which implies that $\underset{n \to \infty}{lim}||\hat{\mX} \hat{\mX}^T - \mX \mX^{T}|| =0$.
\end{proof}

In general, the assumption $O(\|\hat{\mX}_i\|), O(\|\mX_i\|) = O(\mathrm{polylog}(n))$ only holds under certain conditions. 

\subsection{Alternative Embedding Definition with ASE}\label{alt_Y_ASE}

For an adjacency matrix $\mA \in \{0,1\}^{n \times n}$ which represents a binary $n$ node network, typically when this is embedded spectrally via a truncated SVD, it is as:
\begin{equation}
    \mA \approx \mU_{\mA} \mSigma_{\mA} \mV_{\mA}^T,
\end{equation}
where the largest $d\ll n$ singular values are kept. 
Here, $\boldsymbol{\Sigma}_\bmgA \in \mathbb{R}^{d \times d}$ is a diagonal matrix of the $d$ largest singular values of $\bmgA$ in descending order, with $\mU_\bmgA \in \mathbb{R}^{n \times d}$ and $\mV_\bmgA \in \mathbb{R}^{(B+1)n \times d}$ matrices containing, as columns, corresponding orthonormal left and right singular vectors respectively. 

The right spectral embedding is then given by $\hat{\mY} 
= \mV_{\mA}|\mSigma_{\mA}|^{\frac{1}{2}} 
\in \sR^{n \times d}$ and the left spectral embedding by $\hat{\mX}_\mA = \mU_\mA |\mSigma_\mA|^{\frac{1}{2}} \in \sR^{n \times d}$. 
However, we could define an alternative embedding as $\hat{\mY}_{alt} 
= \mV_\mA \mSigma_\mA 
=  \hat{\mY}_\mA |\mSigma_\mA|^{\frac{1}{2}}
\in \sR^{n \times d}$. 

The values of the diagonal matrix $\mSigma_\mA \in\sR^{d \times d}$ are in descending order, and give information about how much variance is explained by each dimension of the variance, as in Principle Component Analysis (PCA) \citep{pearson1901PCA}. 
Thus by using $\hat{\mY}_{alt}$ in Algorithm \ref{alg:knn_bootstrap} when using a spectral embedding to locate each node's $k$-nearest neighbours, the variance captured by each dimension is better represented, and the algorithm is less sensitive to the choice of $d$.

\subsection{4 Community Stochastic Block Model Example}\label{app:4commSBMex}

For simulated data, we generate a $C=4$ community symmetric Stochastic Block Model (SBM) \citep{holland1983stochastic}, an example of a BIRG, of $n=1000$ nodes, where nodes are assigned to each community with equal random chance. The communities can be stored in a community allocation vector $\vtau \in \{1,...,C\}^n$. 
The probability of an edge between two nodes depends on their respective community memberships. 
Specifically, we sampled from an SBM defined by the block probability matrix $\mB \in [0,1]^{C \times C}$, where $\mB$ is given by: 
\begin{equation}\label{Bmatrix_4commSBM}
    \mB = \begin{bmatrix}
0.7 & 0.4 & 0.2 & 0.5 \\
0.4 & 0.6 & 0.3 & 0.2 \\
0.2 & 0.3 & 0.8 & 0.4 \\
0.5 & 0.2 & 0.4 & 0.9 \\
\end{bmatrix}.
\end{equation}

An adjacency matrix $\mA \in \{0,1\}^{n \times n}$ is sampled as: 
\begin{equation}\label{eq:4commSBM}
    \mA_{ij} \overset{\text{ind}}{\sim} \text{Bernoulli}(\mB_{\tau_i, \tau_j}),
\end{equation}
where $\tau_i, \tau_j \in \{1,...,C\}$ denote the communities that nodes $i$ and $j$ belong to respectively. 
A block model $\mB$ is specified the same as for a MMSBM (see Section \ref{section:synthetic_data_example}). 
By defining $\mB$ to be symmetric, $\mB_{g,h}=\mB_{h,g}$ for all $g,h\in\{1,...,C\}$, we have that $\mathbb{P}(\mA_{ij}) = \mathbb{P}(\mA_{ji})$ for all $i,j \in \{1,...,n\}$, however this does not guarantee that $\mA_{ij}=\mA_{ji}$. To ensure symmetric samples, whenever $i>j$, i.e. $\mA_{ij}$ falls into the lower triangle of $\mA$, then set $\mA_{ij} = \mA_{ji}$ to ensure symmetry. 
In this way, we ensure that the resulting adjacency matrix $\mA$ is symmetric. 

\begin{figure}[!htb]
    \centering
    \includegraphics[width=\linewidth]{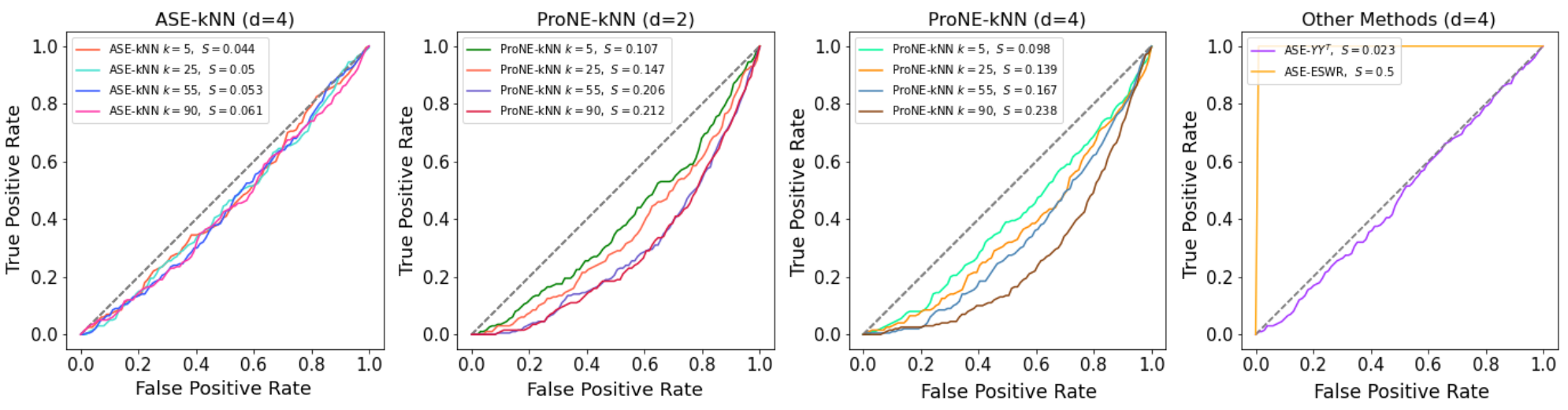}
\caption{Simulated results for the $n=1000$ node 4 community SBM. Each curve is found using $M=200$ draws from the known model, each paired with a single bootstrap from the model. The area between the curve and the $x=y$ line is the Bootstrap Validity Score. }
\label{table:4commSBMex}
\end{figure}

As introduced in Section \ref{section:bootstrap_procedure_validation}, we can use Algorithm \ref{alg:test} $M \in \mathbb{N}$ times to verify if a procedure for generating bootstrapped networks creates exchangeable networks. 
For this example we take $M=200$ random samples from the model specified in Equation \ref{eq:4commSBM} to create the adjacency matrices $\mA^{(1)},...,\mA^{(M)}$. 
For each matrix one bootstrap replicate is created as in Algorithm \ref{alg:knn_bootstrap} with $k=5$ chosen, given by $\widetilde{\mA}^{(1)},...,\widetilde{\mA}^{(M)}$, respectively. 
To each pair $[\mA^{(m)}, \widetilde{\mA}^{(m)}]$, for $m=1,...,M$, we apply Algorithm \ref{alg:test}, and use the $M$ $\hat{p}$-values to create a QQ-plot. 
We repeat the above kNN-based bootstrap (as in Section \ref{section:knnBootstrap}) and change $k$ to be $k=25$ and $k=240$, as well as applying other bootstrap methods (see Appendix \ref{appendix:used_bootstrap_methods} for more details), with different embedding dimension $d$. 
In Figure \ref{table:4commSBMex} we show the QQ-plots and Bootstrap Validity Scores produced by each bootstrapping method.

In Appendix \ref{appendix:knn_sensitivity_k}, we perform a sensitivity analysis to demonstrate the procedure is robust to the choice of $k$ for ASE-kNN, and principled in how it behaves.

\subsubsection{Sensitivity of kNN-bootstrap Procedure to $k$}\label{appendix:knn_sensitivity_k}

\begin{figure}[htbp!]
    \centering
    \includegraphics[width=0.6\linewidth]{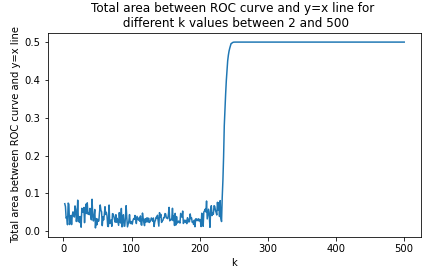}
    \caption{Plot showing possible choices of $k$ against the `Bootstrap Validity Score' (area between the curve and the $x=y$ line) for a 4 community SBM model of $n=1000$ nodes. The smaller the score value, the better the bootstrap method performs at producing exchangeable networks. 
    From the plot we see that choices of $k<200$ all have reasonable score values. 
    Around the $k=230$ mark the score values increase rapidly and plateau at a Bootstrap Validity Score of 0.5. 
    As each of the $n=1000$ nodes are equally likely to belong to each of the 4 communities, the expected size of each community is 250. The kNN-based algorithm performs poorly when $k$ is chosen larger than the size of the smallest community, but this is sensible, as a node is not well represented by a node outside of its own community in this SBM setting. }
    \label{fig:4commSBM_k_scores}
\end{figure}

To demonstrate how the choice of $k$ influences the performance of the kNN algorithm, we uses the $M=200$ draws given by $\mA^{(1)},...,\mA^{(M)}$ from the $n=1000$ node 4 community SBM with block probability matrix $\mB$ given in Equation \ref{Bmatrix_4commSBM} introduced in Section \ref{app:4commSBMex}. 
For all values of $k \in \{2,...,\frac{n}{2}\}$, each of the $M$ matrices are bootstrapped once as in Algorithm \ref{alg:knn_bootstrap} (ASE-kNN) with $d=4$, giving $M$ pairs for each $k$ value. To each pair $[\mA^{(m)}, \widetilde{\mA}^{(m)}]$, for $m = 1,...,M$, we apply Algorithm \ref{alg:test}, and use the $M$ $p$-values to create a QQ-plot. 
The Bootstrap Validity Score is calculated for each value of $k \in \{2,...,\frac{n}{2}\}$, and gives a measure of how well the bootstrap algorithm performs with this data example for different choices of $k$. 
In Figure \ref{fig:4commSBM_k_scores} the choices of $k$ are plotted against their scores. 
By observing the score values, we see that there is a wide range of choices of $k$ that the algorithm performs well for with this data. 
Since in this example we know that $n=1000$ for 4 communities, where all communities have equal probability of a node belonging to it (i.e. the expected size of each community is 250), we see the algorithm begins to perform poorly when $k$ is chosen to be larger than the size of the smallest community in an observed network. 

\newpage
\subsection{Statement of Considered Bootstrap Algorithms}\label{appendix:used_bootstrap_methods}

\begin{algorithm}
\caption{$\mX\mX^T$ Bootstrap of an Unweighted Network}\label{alg:XYT_bootstrap}
\begin{algorithmic}[1]
\State \textbf{Input:}
\State \quad Observed adjacency matrix $\mA \in \{0,1\}^{n \times n}$
\State \quad Embedding dimension $d \leq n$
\State \quad Number of bootstrapped graphs $B$
\State Compute the $d$-dimensional adjacency spectral embedding  $\hat{\mX}_\mA$ of $\mA$
\State Set $\hat{\mP} = \hat{\mX}_\mA \hat{\mX}_\mA^T$
\State Set any values in $\hat{\mP}<0$ to be 0, and set any values in $\hat{\mP}>1$ to be 1
\For {$b=1,\dots, B$}
\State Sample \[\widetilde{\mA}^{(b)}_{ij} \overset{\text{ind}}{\sim} \text{Bernoulli}(\hat{\mP}_{ij})\]
\EndFor
\State \textbf{Output:} $\widetilde{\mA}^{(1)},\dots,\widetilde{\mA}^{(B)}$
\end{algorithmic}
\end{algorithm}

A na\"{i}ve way to bootstrap a network would be to consider the edge list $E$ of an adjacency matrix $\mA \in \{0,1\}^{n \times n}$, and sample $|E|$ edges with replacement. As we are considering binary graphs, any edge sampled more than once will have it's duplicates removed from the final sample. The resulting edge list $\widetilde{E}$ will populate an adjacency matrix $\widetilde{\mA}\in\{0,1\}^{n\times n}$. This is given by Algorithm \ref{alg:edgelistsample_bootstrap}. 

\begin{algorithm}
\caption{Edge List Sample with Replacement (ESWR) Bootstrap of an Unweighted Network}\label{alg:edgelistsample_bootstrap}
\begin{algorithmic}[1]
\State \textbf{Input:}
\State \quad Observed adjacency matrix $\mA \in \{0,1\}^{n \times n}$
\State \quad Number of bootstrapped graphs $B$
\State Create the edge list $E$ of edges in $\mA$
\For {$b=1,\dots, B$}
\State Sample with replacement $|E|$ edges from $E$ to give an edge list ${\widetilde{E}}^{(b)}$
\State Update $\widetilde{E}^{(b)}$ by removing any duplicate edges
\State Use $\widetilde{E}^{(b)}$ to construct a binary adjacency matrix $\widetilde{A}^{(b)}$
\EndFor
\State \textbf{Output:} $\widetilde{\mA}^{(1)},\dots,\widetilde{\mA}^{(B)}$
\end{algorithmic}
\end{algorithm}


We can extend Algorithm \ref{alg:edgelistsample_bootstrap}, by adding random edges to the bootstrapped $\widetilde{\mA}^{(b)}$ for $b=1,...,B$ outputted by Algorithm \ref{alg:edgelistsample_bootstrap}, such that all bootstrap resamples have the same number of edges as the observed adjacency matrix $\mA$. See Algorithm \ref{alg:edgelistsample_bootstrap_randomEdges}. 

\begin{algorithm}\label{alg:ESWRplusEdges}
\caption{Edge List Sample with Replacement Bootstrap + Random Edges of an Unweighted Network}\label{alg:edgelistsample_bootstrap_randomEdges}
\begin{algorithmic}[1]
\State \textbf{Input:}
\State \quad Observed adjacency matrix $\boldsymbol{A} \in \{0,1\}^{n \times n}$
\State \quad Number of bootstrapped graphs $B$
\State Create the edge list $E$ of edges in $\mA$
\For {$b=1,\dots, B$}
\State Sample with replacement $|E|$ edges from $E$ to give an edge list ${\widetilde{E}}^{(b)}$
\State Update $\widetilde{E}^{(b)}$ by removing any duplicate edges
\State Find $a \in \mathbb{N}_0$ such that $|\widetilde{E}^{(b)}|+a=|E|$
\State Sample without replacement $a$ edges from $(\widetilde{E}^{(b)})^c$ and add these to $\widetilde{E}^{(b)}$
\State Use $\widetilde{E}^{(b)}$ to construct a binary adjacency matrix $\widetilde{\boldsymbol{A}}^{(b)}$
\EndFor
\State \textbf{Output:} $\widetilde{\boldsymbol{A}}^{(1)},\dots,\widetilde{\boldsymbol{A}}^{(B)}$
\end{algorithmic}
\end{algorithm}

\FloatBarrier




\subsection{MMSBM - effects of changing n to the Bootstrap Validity Score with Different Methods}\label{app:MMSBM_n}

\begin{figure}[htbp!]
    \centering
    \includegraphics[width=0.85\linewidth]{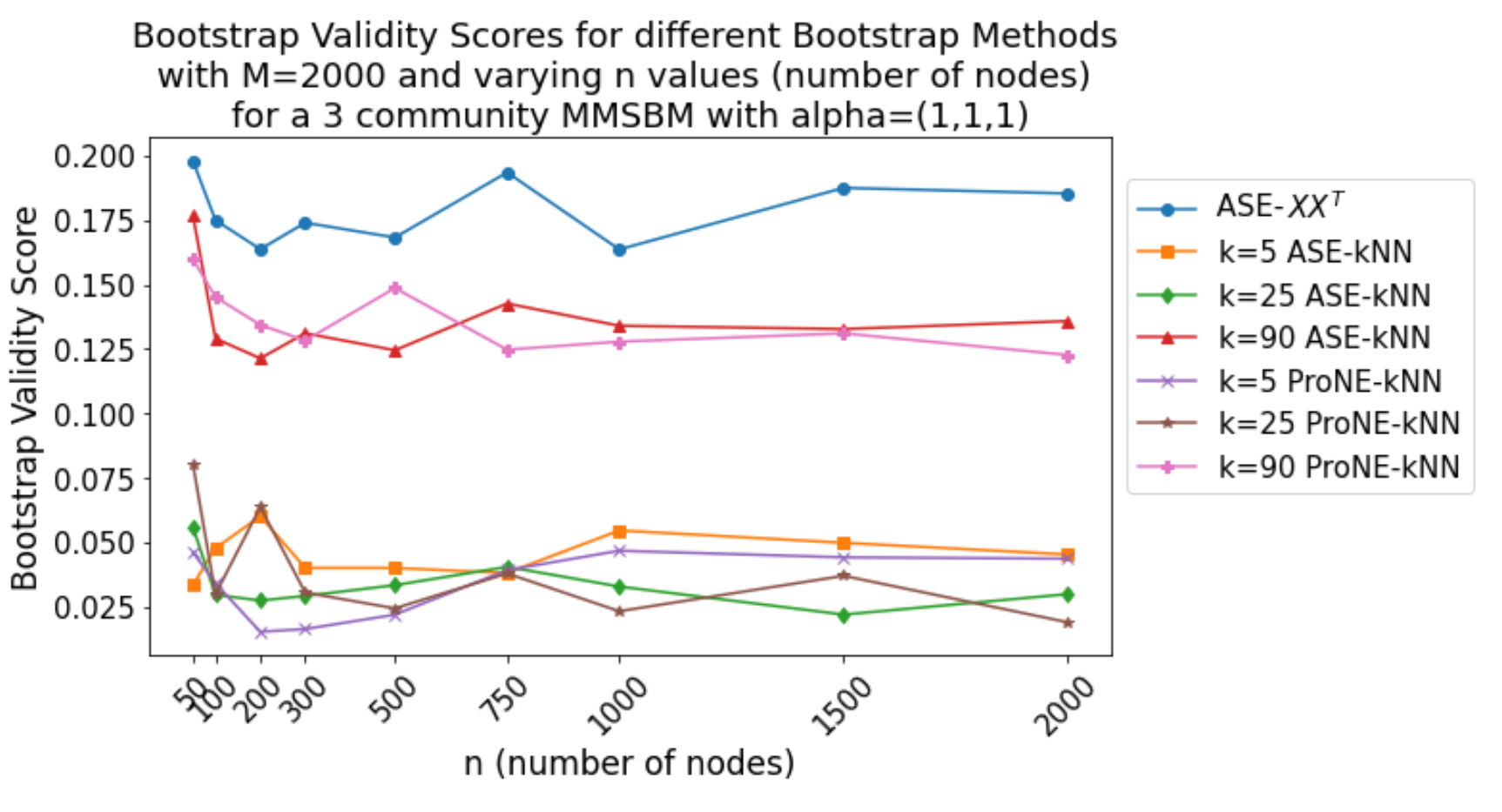}
    \caption{MMSBM, $C=3$ communities, with $\bm\alpha = \bm{1}_C$ for different values of $n$ and different embedding methods and bootstrap procedures applied.}
    \label{fig:MMSBM_n}
\end{figure}

We look to see how the network size $n$ changes the Bootstrap Validity Score for different methods in Figure \ref{fig:MMSBM_n}. 
We see for this synthetic data example, ASE-$XX^T$ performs worse than all other methods we tried. We apply the kNN bootstrap procedure with ASE into $d=3$ dimensions (as in Algorithm \ref{alg:knn_bootstrap}) and use ProNE embedding (into $d=2$ dimensions as Figure \ref{fig:MMSBMex} showed $d=2$ performed better than $d=3$ with ProNE) with kNN bootstrapping. 
For the same values of $k$, the ASE-kNN and ProNE-kNN perform 
fairly similarly across all values of $n$, as shown by Figure \ref{fig:MMSBM_n}. We see for $k=90$, both methods yield a fairly high Bootstrap Validity Score, likely that $k$ is chosen too large. For $k=5$ and $k=25$ we see fairly similar performance, however when $n>750$, it appears that $k=25$ is a better choice than $k=5$ for both embedding method choices. 
 
\subsection{Overview of t-SNE}\label{tsneOverview}

t-SNE (t-distributed Stochastic Neighbour Embedding) was introduced in \cite{van2008visualizing} as a tool for visualising high-dimensional data in two or three dimensions. 
In the original high-dimensional space, pairwise similarities between data points are calculated using a probability distribution, which preserves local structures by assigning higher probabilities to points that are close together. 
t-SNE tries to replicate the high-dimensional similarities in a low-dimension space by optimising the Kullback-Leibler (KL) divergence \citep{kullback1951information} between the high-dimension and low-dimension probability distributions, using gradient descent. 

t-SNE is good at clustering and separating groups with different local relationship structure. However, it does not preserve global structures well.
The algorithm can separate clusters well, but where the clusters are embedded in the low-dimensional t-SNE space is not necessarily representative of how similar certain clusters are in behaviour, or where there are across cluster local relationships. 
The perplexity hyperparameter, which controls the balance between local and global aspects of the data, thus needs tuning carefully. It is also of note that t-SNE is non-deterministic, so results may vary across runs. A fixed seed can be used to create reproducible t-SNE embeddings.

\subsection{Sensitivity Analysis Checks of embedding of the School Data Example}\label{robustnessChecks_tSNE_school}

\begin{figure}[htbp!]
    \centering
    \includegraphics[width=0.56\linewidth]{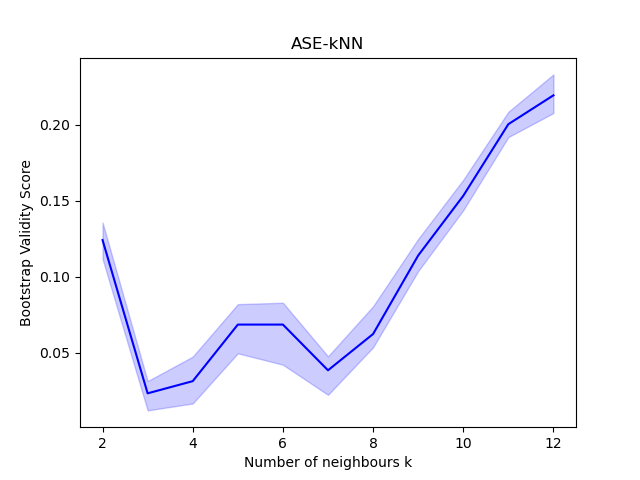}
    \caption{Bootstrap Validity Score for the School Data averaged over 20 runs, with 90\% confidence intervals, for $k \in \{2,...,12\}$. }
    \label{fig:knn_choosek}
\end{figure}

\begin{figure}
    \centering
    \includegraphics[width=\linewidth]{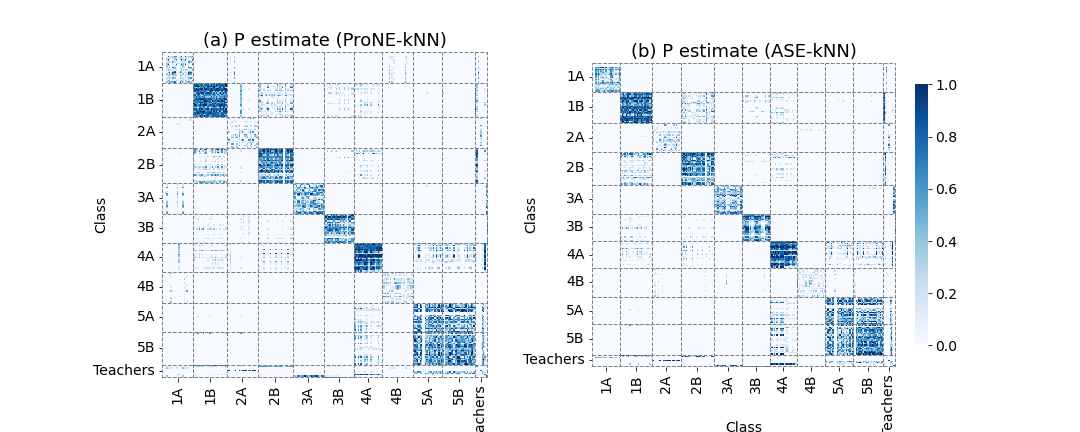}
    \caption{$\hat{\mP}$ for the School Data ASE-kNN ($d=10$) or ProNE-kNN ($d=6$) embedding (both $k=5$). Both are very similar, having passed the Bootstrap Exchangeability Test.}
    \label{fig:phat}
\end{figure}

To choose $k$ in the kNN model, we computed the Bootstrap Validity Score for each $k \in \{2,...,12\}$. We find that $k$ being too small or too large leads to poor bootstraps, but in between there is tolerance between $k=3$ and $8$ (Figure \ref{fig:knn_choosek}). So as not to overfit, we choose the middle value $k=5$, even though $k=3$ or $7$ are possibly better. For both ProNE and ASE, this leads to very similar $\hat{\mP}$ values (Figure \ref{fig:phat}).

\begin{figure}
    \centering
    \includegraphics[width=0.9\linewidth]{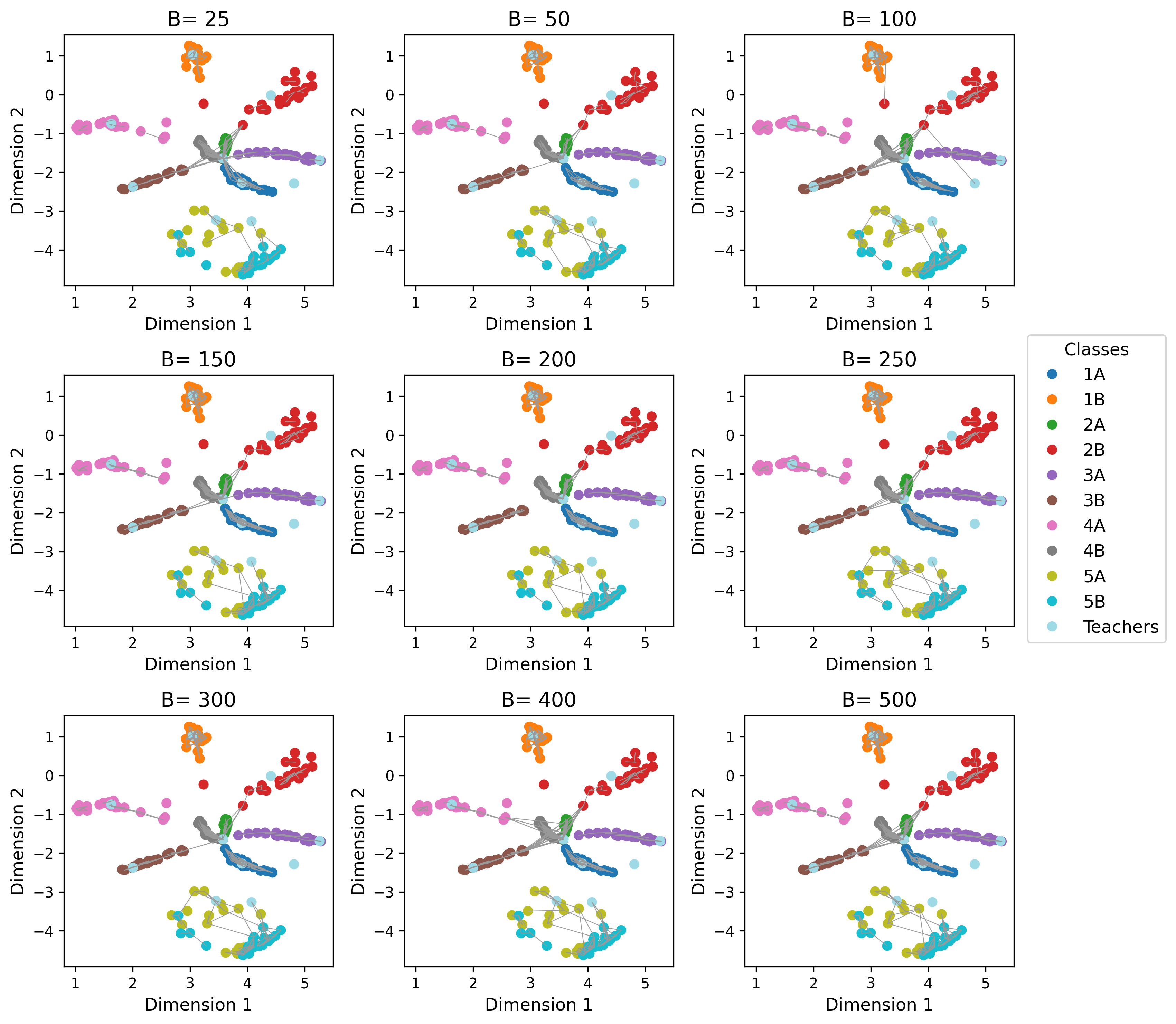}
    \caption{Sensitivity to $B$ shown on a shared t-SNE visualisation applied to the observed adjacency matrix with Perplexity 125, $\mA^{(obs)}$ is shown. The edges connecting nodes are from within 3 SDs of one another in the $d$ dimensional embedding space using ASE-kNN ($k=5$, $d=10$, varied $B$).}
    \label{fig:tsne_sensitivity_B}
\end{figure}

When estimating the covariance of each node in the $d$ dimensional embedding space created via UASE, we have $B$ bootstrap embeddings and 1 observed embedding to compute the estimate. Here we show that the method is not sensitive to the number of bootstraps used, by plotting the t-SNE embedding of $\mA^{(obs)}$, with the t-SNE perplexity value fixed at perplexity = 125. We set $B = 25, 50, 100, 150, 200, 250, 300, 400, 500$ in our sensitivity analysis. 
We set $d=10$ as the ASE embedding dimension and $k=5$ for the kNN to compute $\hat
{\mP}$. 
The t-SNE embeddings are standardised to be mean 0, standard deviation 1. Figure \ref{fig:tsne_sensitivity_B} shows that the method is robust to changing the value of $B$, the number of bootstraps, as the lines between clusters are stable. For each value of $B$, the ``fuzziness'' matrix $\mF$ is calculated, and a line is plotted between nodes $i$ and $j$ when $\mF_{ij}=1$. We have $\mF_{ij}=1$ when the embeddings of nodes $i$ and $j$ in the $d$ dimensional embedding space are both within 3 standard deviations of one another, using the $B$ bootstrap embeddings alongside the observed network's embedding to estimate the covariance of each node.

\begin{figure}
    \centering
    \includegraphics[width=0.6\linewidth]{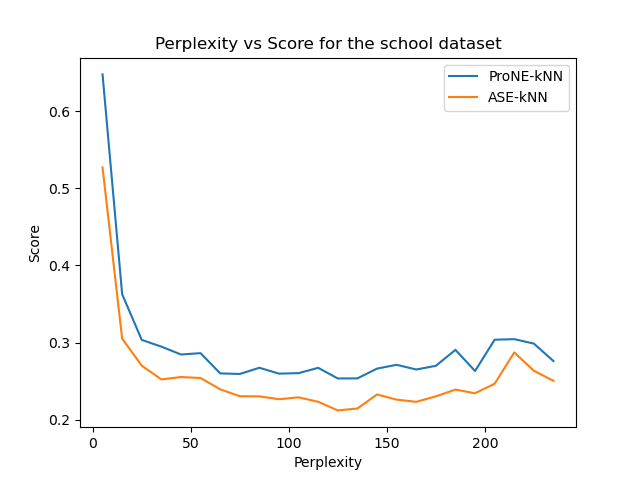}
    \caption{Fuzziness Score as a function of t-SNE Perplexity for the School Data ASE-kNN ($d=10$) or ProNE-kNN ($d=6$) embedding (both $k=5$). Both methods agree  that `good' values are around 125 with a wide tolerance of $\pm$75.}
    \label{fig:tsne_perplexity}
\end{figure}

\begin{figure}[!htbp]
    \centering
    \includegraphics[width=0.9\linewidth]{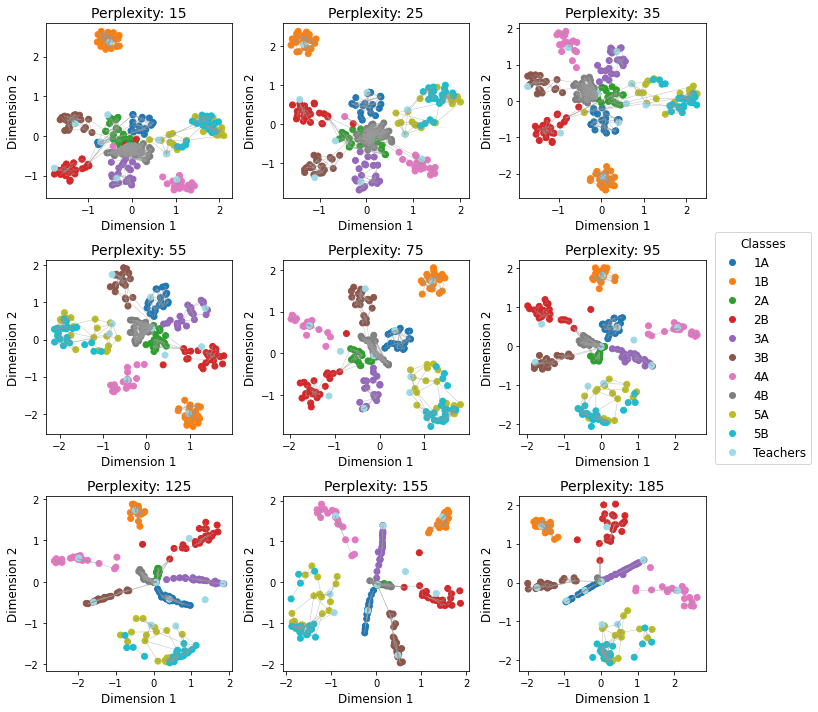}
    \caption{Plots of t-SNE applied to the observed adjacency matrix $\mA^{(obs)}$ for different perplexity values. For $B=500$ bootstraps the ``fuzziness'' matrix $\mF$ is calculated, and a line is plotted between nodes $i$ and $j$ if $\mF_{ij}=1$, i.e. nodes $i$ and $j$ have overlapping distributions. We see the method is robust to different choices of perplexity. }
    \label{fig:tsne_sensitivity_perplexity}
\end{figure}

The t-SNE algorithm has a parameter called perplexity, which can be interpreted as a smooth measure of the effective number of neighbours used by the algorithm to calculate a nodes similarity to other nodes \citep{van2008visualizing}. For a fixed number of bootstraps, $B=500$, we calculate the covariance of each node, and thus the ``fuzziness'' matrix $\mF$. 
A systematic scan of perplexity is shown in Figure \ref{fig:tsne_perplexity}, summarised with the Fuzziness score. We then visualise the corresponding t-SNE of $\mA^{(obs)}$ for perplexity $=15, 25, 35, 55, 75, 95, 125, 155, 185$, with edges from $\mF$, in Figure \ref{fig:tsne_sensitivity_perplexity}. The method is robust to perplexity choice, however when perplexity is chosen to be larger than the largest community size, the t-SNE algorithm will certainly explore relationships outside of the cluster each node belongs to, which is valuable. 

For the school data example, each class had between 22 and 26 students, with one teacher per class \citep{stehle2011high}. The largest cluster size is therefore 27, if we consider the largest class and their teacher. When perplexity is increased to a value larger than 27, we see that the t-SNE algorithm does a better job at separating nodes, especially near the origin. In Figure \ref{fig:tSNE_school} we can see that when perplexity is increased from 15 to 125 the algorithm does a much better job at clustering the data, as it is able to look outside of the members of the cluster to explore the global data structure, yielding a better visualisation.

\begin{figure}
    \centering
    \includegraphics[width=\linewidth]{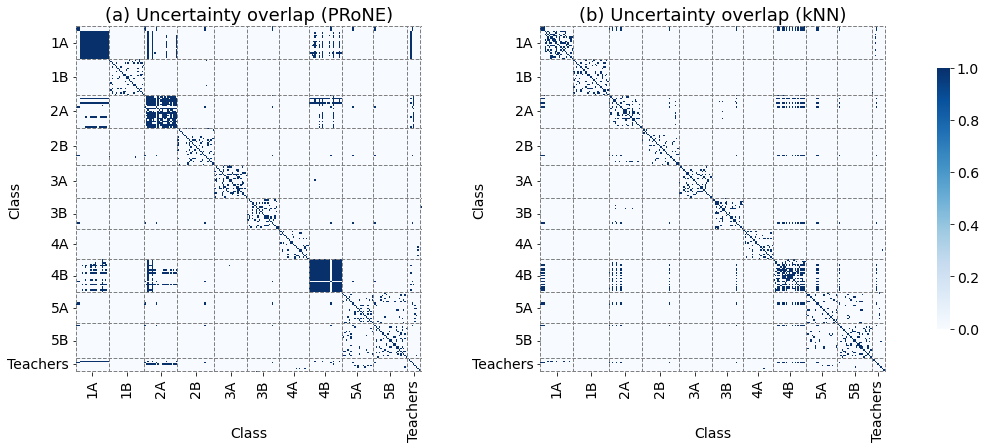}
    \caption{Visualisation of the ``fuzziness'' matrix $\mF$ calculated using $B=500$ bootstraps, for ProNE-kNN ($d=6$) and ASE-kNN ($d=10$). Where $\mF_{ij}=1$, nodes $i$ and $j$ are within 3 SDs of one another, i.e. the nodes have overlapping distributions. Both methods highlight the same within cluster and across cluster structures. }
    \label{fig:FigFriendsProneVsKnn}
\end{figure}

Our uncertainty estimates are relatively robust to the choice of $\mP$ estimate. Figure \ref{fig:FigFriendsProneVsKnn} shows the inferred $\mF$ matrices using the two best bootstraps, either ProNE-kNN (d=6) or ASE-kNN (d=10). Whilst differences exist, these are within-cluster structures and the important cross-cluster features are present similarly in both.

\end{document}